\documentclass[a4paper,fleqn,usenatbib]{mn2e}
\usepackage{graphicx}
\usepackage{amssymb}
\usepackage{amsmath}
\usepackage[breaklinks=true,colorlinks,citecolor=blue, linkcolor=black, urlcolor=blue]{hyperref}
\usepackage[pass]{geometry}
\usepackage[T1]{fontenc}
\usepackage{enumerate}
\usepackage{enumitem}
\usepackage{color}
\usepackage{units}
\usepackage{natbib}
\usepackage[draft]{varioref}
\usepackage{adjustbox}
\usepackage{subcaption}
\usepackage{multirow}
\usepackage{multicol}
\usepackage{csquotes}
\MakeOuterQuote{"}
\definecolor{update}{rgb}{1,0,0}

\defcitealias{Franx1988}{FI88}

\newcommand\blfootnote[1]{
\begingroup
\renewcommand\thefootnote{}\footnote{#1}
\addtocounter{footnote}{-1}
\endgroup
} 

\title[IC~1459 with KMOS and MUSE]{Unravelling the Origin of the Counter-Rotating Core in IC~1459 with KMOS and MUSE}

\author[L. J. Prichard et al.]{Laura J. Prichard$^{1\star,2}$, Sam P. Vaughan$^{2}$, Roger L. Davies$^{2}$\\
$^{1}${\small Space Telescope Science Institute, 3700 San Martin Drive, Baltimore, MD 21218, USA}\\
$^{2}${\small Sub-department of Astrophysics, Department of Physics, University of Oxford, Denys Wilkinson Building, Keble Road, Oxford OX1 3RH, UK}\\
}

\date{Accepted 2019 April 24. Received 2019 April 18; in original form 2018 July 17}
\pubyear{2019}

\begin{document}
\label{firstpage}
\maketitle

\begin{abstract}
The massive early-type galaxy (ETG) IC~1459 is a slowly rotating galaxy that exhibits a rapidly counter-rotating kinematically decoupled core (KDC, $R_{\rm KDC}\approx 5^{\prime\prime}\approx 0.1 R_{\rm e}$). To investigate the origin of its KDC, we coupled large data mosaics from the near-infrared (NIR)/optical integral field unit (IFU) instruments K-band Multi-Object Spectrograph (KMOS) and Multi Unit Spectroscopic Explorer (MUSE). We studied IC~1459's stellar populations and, for the first time for a KDC, the spatially resolved initial mass function (IMF). We used full-spectral-fitting to fit the stellar populations and IMF simultaneously, and an alternative spectral-fitting method that does not assume a star-formation history (SFH; although does not constrain the IMF) for comparison. When no SFH is assumed, we derived a negative metallicity gradient for IC~1459 that could be driven by a distinct metal-poor population in the outer regions of the galaxy, and a radially constant old stellar age. We found a radially constant bottom-heavy IMF out to $\sim\nicefrac{1}{3}R_{\rm e}$. The radially flat IMF and age extend beyond the counter-rotating core. We detected high velocity dispersion along the galaxy's major axis. Our results potentially add weight to findings from orbital modelling of other KDCs that the core is not a distinct population of stars but in fact two smooth co-spatial counter-rotating populations. No clear picture of formation explains the observational results of IC~1459, but we propose it could have included a gas-rich intense period of star formation at early times, perhaps with counter-rotating accreting cold streams, followed by dry and gas-rich mergers through to the present day.
\end{abstract}


\section{Introduction}
\label{sec:intro}
\blfootnote{\vspace{-4 mm}$^{\star}$Email: \href{mailto:lprichard@stsci.edu}{lprichard$@$stsci.edu}}

Large integral field unit (IFU) surveys of local galaxies have revolutionised our understanding of early-type galaxies \citep[ETGs; see][for a review]{Cappellari2016}. Spatially resolved kinematics revealed two distinct classes of ETG \citep[e.g.,][]{deZeeuw2002, Cappellari2011a}. The two classes can generally be separated by mass, where the irregularly rotating galaxies dominate above $M_{\mathrm{crit}}~\approx~2~\times~10^{11}\; \mathrm{M}_{\odot}$ \citep[i.e. `slow rotators';][]{Emsellem2004, Cappellari2007, Emsellem2007}. The typically lower-mass regular/fast rotators make up the bulk of ETGs \citep[$\sim85\%$;][]{Emsellem2007, Emsellem2011} and have simple disc-like velocity maps \citep{Krajnovic2008, Krajnovic2011}.  

The ATLAS$^{\text{3D}}$ survey \citep{Cappellari2011a} led to further sub-classification of the slow rotators \citep{Krajnovic2011}, one of which is galaxies that exhibit kinematically distinct/decoupled cores (KDCs). KDCs are relatively rare in ETGs as a whole ($\sim7\%$) but are common among slow rotators \citep[$\sim42\%$;][]{Krajnovic2011}. Studies of the stellar populations of galaxies with KDCs have shown old ages and negative metallicity gradients (e.g., \citealt{Efstathiou1985, Franx1988}, hereafter \citetalias{Franx1988}; \citealt{Gorgas1990,Bender1992a,Carollo1997b, Carollo1997a,Mehlert1998,Davies2001,Emsellem2004,Kuntschner2010,McDermid2015}). This is similar to ETGs of comparable masses with no KDC. The light profiles of galaxies with KDCs \citep[e.g.,][]{Efstathiou1980, Lasker2014} are also similar to ETGs without them \citep[e.g.,][]{Forbes1995, Carollo1997b, Carollo1997a, Emsellem2011, Lauer2012, Krajnovic2013}. 

There is a wealth of evidence that supports a two-stage formation of massive slow-rotator ETGs \citep[e.g.,][]{Oser2010}. The first phase is thought to be dominated by rigorous star formation (i.e. `starburst') that could be triggered by gas-rich mergers \citep[e.g.,][]{Barnes1991} or rapid accretion of cold streams \citep[e.g.,][]{Dekel2009a, Elmegreen2010, Dekel2014, Forbes2014}. The second phase of evolution is thought to be the steady accretion of stars from gas-poor satellites \citep[e.g.,][]{Naab2007, Naab2009, Hopkins2009, Feldmann2011, Johansson2012, Lackner2012}. It was originally thought that KDCs had an external origin and provided conclusive proof of major mergers to form ETGs (\citealt{Efstathiou1980}; \citetalias{Franx1988}; \citealt{Bender1992a}). Counter rotation could also arise from the accretion of gas or gas rich satellites (e.g., \citealt{Bertola1988}; \citetalias{Franx1988}). However, these gas-rich evolutionary scenarios are in conflict with the standard definition of the two-phase path of ETG formation and are therefore difficult to reconcile with the observational similarities of KDC galaxies with other massive ETGs. 

Many studies have shown that massive galaxies with KDCs show radially constant ages (e.g., \citetalias{Franx1988}; \citealt{Bender1988, Rix1992, Davies2001, McDermid2006b, Kuntschner2010, McDermid2015, Krajnovic2015}). These findings support an interesting development in the study of KDCs from dynamical orbital modelling \citep{vandenBosch2008, Krajnovic2015}. \cite{vandenBosch2008} showed that when modelling the galaxy NGC 4365, its KDC was not an orbitally distinct component but actually an observational effect of two smoothly distributed prograde and retrograde populations of stars (explaining the radially constant age) on thick short-axis tube orbits \citep[as confirmed by][]{Krajnovic2015}. The studies found that the net rotation of the KDC arises from the mass imbalance of stars on the prograde and retrograde orbits. Despite this advance in our understanding of KDCs, it is not clear what formation mechanisms would result in prograde and retrograde populations of stars.

The initial mass function (IMF) is the mass distribution of a newly formed population of stars. Radial gradients in the IMF within a galaxy can provide information about its possible evolutionary path. For example, it is thought that a bottom-heavy IMF towards the centres of massive ellipticals supports the two-phase growth channel where the cores are formed first and extremely rapidly \citep[e.g.,][]{Chabrier2014, vanDokkum2015, Zolotov2015, Barro2016}. So far only a handful of studies have investigated spatial variation of the IMF within galaxies \citep{Martin-Navarro2015a, Martin-Navarro2015b, Zieleniewski2015, Zieleniewski2017, McConnell2016, LaBarbera2016, vanDokkum2017, Parikh2018, Sarzi2018, Vaughan2018a, Vaughan2018b}. As yet, radial measurements of the IMF in a galaxy with a KDC have not been made and could provide valuable clues as to their formation.

To help understand the formation of galaxies that exhibit KDCs, we investigated IC~1459; the archetypal massive ETG with a counter-rotating core \citepalias{Franx1988}. IC~1459 is a bright \citep[$M_V\simeq-22.3$; Lyon-Meudon Extragalactic Database -- LEDA, see][]{Paturel1997}, local \citep[D~$=30.3\pm4.0$~Mpc;][]{Ferrarese2000}, massive \citep[$M\sim4\text{--}6\times10^{11}\;\mathrm{M}_{\odot}$;][]{Cappellari2002, Samurovic2005} E3 ETG. It is the central galaxy in the gas-rich IC~1459 group of 11 galaxies that is dominated by spirals \citep[group number 15;][]{Huchra1982, Serra2015}. The outer parts of the galaxy and ionized gas in the centre show rotation in one direction, while the central stellar component is rapidly counter rotating with a maximum velocity $\sim170 \pm 20$~km~s$^{-1}$ \citepalias{Franx1988}.

To investigate the evolution of IC~1459, we coupled large field-of-view (FOV) near-infrared (NIR) data from the K-band Multi-Object Spectrograph \citep[KMOS;][]{Sharples2013} and optical data from the Multi Unit Spectroscopic Explorer \citep[MUSE;][]{Bacon2010}. Both instruments are on the Very Large Telescope (VLT) at Cerro Paranal in Chile. In this paper we give information on the KMOS data (Section \ref{sec:kmosic1459}) and MUSE data (Section \ref{sec:muse}) we used to investigate the properties of IC~1459. We then describe the analysis of the mosaics of data to measure the kinematics (Section \ref{sec:moskin}), stellar populations and IMF (Section \ref{sec:icstellpops}) of IC~1459 in order to understand its evolution. We discuss the possible implications of this analysis in Section \ref{sec:icdisc} and summarise the conclusions of this work in Section \ref{sec:icconc}. Throughout this paper, we assume $\Lambda$ Cold Dark Matter cosmology with $\Omega_m = 0.3$, $\Omega_{\Lambda} = 0.7$, $H_0 = 70$ km s$^{-1}$ Mpc$^{-1}$ (these agree well with the latest results from \citealt{Planck2016}) and use the AB magnitude system \citep{Oke1983}.

\section{Mosaic Data}
\label{sec:mosdat}

\subsection{KMOS Data}
\label{sec:kmosic1459}

\subsubsection{KMOS Observations}
\label{sec:kmosicobs}

IC~1459 has an angular size $R_{\mathrm{e}}=46.2\arcsec$ \citep{Lasker2014}. Due to the size and magnitude of the galaxy, IC~1459 makes the ideal backup target for variable weather conditions or poor seeing. KMOS is a multi-object, integral-field, infrared spectroscopic instrument. The 24 configurable arms of KMOS move within a $7.2\arcmin$ diameter patrol field. It would take $\sim 16.5\times 16.5$ KMOS IFUs (each $2.8\arcsec \times 2.8\arcsec$, with $14\times 14$ $0\farcs2$ pixels) to span the half-light radius of IC~1459. In order to cover a sufficient area, one can use KMOS's highly under-utilised `Mosaic Mode'. Here the 24 arms of KMOS can be arranged in a grid of fixed dimensions and moved across the surface of a target to create a complete mosaic of data. In the `mapping24' mode, the arms are arranged in a $6\times 4$ grid and with 16 successive pointings (with a $~1$~pixel overlap) can cover a contiguous region of $\sim64.9\arcsec \times 43.3\arcsec$.

The KMOS data for IC~1459 were taken in two different observing periods. The useable observations across both periods were taken in two observing blocks (OBs) in the first period and one OB in the second period. For simplicity, we will call these Mosaics 1a and 1b (taken in P93) and Mosaic 2 (taken in P95). Mosaics 1a and 1b were observed on 10$^{\mathrm{th}}$ July 2014 in P93 as a poor seeing ($\sim1.5\text{--}2\arcsec$) backup target for the KMOS Clusters Survey \citep[during ESO programme ID: 093.A-0051(A);][]{Beifiori2017, Prichard2017b, Chan2018}. During this observing period, three of the 24 arms were not functioning. To fill in the gaps in the mosaic from the missing IFUs, the mosaic was observed in two parts rotated by 180$^{\circ}$. The OBs for IC~1459 could not be completed for either position angle, due to the loss of guide-star tracking through thick cloud. IC~1459 was observed again as a backup target in P95 to fill in the gaps of the P93 mosaics. Mosaic 2 was observed on 7$^{\mathrm{th}}$ August 2016 as a poor-seeing ($\sim1\text{--}1.5\arcsec$) backup for the KMOS Redshift One Spectroscopic Survey \citep[during European Southern Observatory -- ESO -- programme ID: 095.B-0035(B);][]{Stott2016}. All the arms were functioning at the time of the observations. Again, due to the variable weather conditions the telescope lost tracking and only a partial mosaic of data could be obtained.

For all three mosaics, the galaxy was observed in the $IZ$ band ($0.8\text{--}1.08 \;\mu \mathrm{m}$, $R\sim3400$ at band centre) with 300~s exposures for every one of the 16 pointings to build a complete mosaic. The mosaic was observed using the standard nod-to-sky mode. All the KMOS observations were taken during dark time.

\subsubsection{KMOS Data Reduction}
\label{sec:icred}

All three mosaics were observed as different OBs and were therefore reduced separately. The data for IC~1459 was reduced using a combination of the ESO KMOS pipeline (\textsc{spark}) and ESO Recipe Execution tool \citep[\textsc{esorex}, version 3.10.2;][]{Davies2013}, and purpose-built software. \textsc{esorex} performs many of the main reduction steps such as dark, flat, wavelength, and illumination corrections and sky subtractions. The noise cubes for the mosaics were not properly propagated in this version of the pipeline. We therefore had to use different methods to estimate the noise when fitting the spectra from the cubes as described in Section \ref{sec:moskin}. We decided not to combine the mosaics at this stage as each required a different telluric correction and it simplified the analysis to treat them separately (as in Section \ref{sec:moskin}). Instead we combined average spectra from the separate mosaics as described in Section \ref{sec:exspec}.

\subsubsection{KMOS Telluric Correction}
\label{sec:tellmos}

To correct for the telluric absorption in the reduced KMOS cubes, we used \textsc{molecfit} \citep{Smette2015, Kausch2015}. The software uses three-hour interval atmospheric data measured at Paranal Observatory from the Global Data Assimilation System (GDAS) to model the absorption by the Earth's atmosphere. Given the length of the KMOS OBs ($\sim1$~hour as only partially completed) and interval length of the GDAS atmospheric information (3~hours), one telluric correction per mosaic was sufficient. We fitted the median spectrum from each mosaic using \textsc{molecfit}. We opted to fit the continuum of the spectra using a high-order polynomial (maximum of eight) which helped to improve the fit to the worst affected telluric absorption regions. We also chose to fit the whole spectral range simultaneously (excluding the very ends where the data was of poorer quality). We found this gave the best fit to the data which we judged by comparing the $\chi^2$ values of each fit. \textsc{molecfit} produced an output best-fitting model telluric spectrum which we divided through every pixel in the cube. This resulted in continuum and telluric corrected cubes for the three KMOS mosaics. 

\subsection{MUSE Data}
\label{sec:muse}

\subsubsection{MUSE Observations}
\label{sec:museobs}

MUSE is a large FOV ($1\arcmin\times1\arcmin$) panchromatic optical ($\sim0.46\text{--}0.93\;\mu \mathrm{m}$) IFU instrument with resolving power $R\sim3000$ at  band centre and a pixel scale of $0.2\arcsec$ (the same as KMOS). When using KMOS in Mosaic Mode, as was done for IC~1459, one is able to cover an area of the sky that is comparable to the MUSE footprint. Given its wide optical spectral coverage, resolving power, FOV and pixel scale, MUSE makes the ideal optical companion to KMOS. 

The MUSE data of IC~1459 were observed on 14$^{\mathrm{th}}$ October 2014 (programme ID 094.B-0298(A), P.I. C. J. Walcher) with seeing $\sim1.3\text{--}1.6\arcsec$. The data were taken with the \texttt{MUSE\_wfmnoao\_obs\_genericoffset} observing template which used a nod-to-sky observing pattern and rotation between exposures to reduce instrumental effects on the output cube. The observations consisted of eight short ($\sim150$ s) object exposures followed by two sky exposures with a $\sim9^\prime$ separation from the galaxy pointing. The observations were dithered by $\lesssim1\arcsec$ to improve the removal of bad pixels (as with KMOS). The released reduced data cube (see Section \ref{sec:museprep}) had a total integration time of 368~s exposure. The MUSE observations were taken during bright time so have increased photon noise.

\subsubsection{MUSE Cube Preparation}
\label{sec:museprep}

Unlike KMOS data, reduced MUSE data can be accessed via the ESO Science Archive\footnote{\href{http://archive.eso.org/cms.html}{\url{http://archive.eso.org/cms.html}}}. The data were reduced using the \textsc{muse-1.6.1} pipeline and were published as a reduced data cube on the 22$^{\mathrm{nd}}$ June 2016. As described in the ESO Phase 3 Data Release Description\footnote{\href{http://www.eso.org/observing/dfo/quality/PHOENIX/MUSE/processing.html}{\url{http://www.eso.org/observing/dfo/quality/PHOENIX/MUSE/processing.html}}}, the pipeline removes most of the observational signatures of the instrument. The method of sky removal used in the reduction recipe \texttt{muse\_scipost} was \texttt{subtract-model}. This recipe performs a model-based sky subtraction -- a model of the sky lines and continuum were computed using the darkest region of the field then the models were subtracted from the data.

In theory, the reduced MUSE data cube of IC~1459 was a science-quality data product that was ready for analysis. However, we found that this was not the case as the data still suffered strong contamination from bright sky lines following the initial sky subtraction. To improve the removal of the sky for the MUSE data of IC~1459, a `pseudo sky' spectrum was taken from the edges of the MUSE cube (where the sky was strong and the contribution of the galaxy was negligible) and was subtracted from the rest of the cube. Given the short exposure of the galaxy and bright sky lines due to being observed in bright time, we found this method sufficient, despite IC~1459 extending to the edges of the cube. This reduction step visibly improved the sky removal in some areas of the cube but residual sky features (from over- and under-subtraction) were still visible across the cube and worst towards the red end of the spectra. This contamination can be seen in the average spectra extracted from the MUSE cube (see Section \ref{sec:exspec}) shown in Figure \ref{fig:icspec}. Despite this residual contamination, we found this method to be the simplest way to get usable data for our science purposes. We used the bluest and least contaminated region with conservative masking of one sky line for kinematics (Section \ref{sec:derivkin}) and stellar populations analysis (see Section \ref{sec:icstellpops}) and we switch to KMOS in the worst affected red region of MUSE from the Ca II triplet feature (CaT, $\lambda8498, 8542, 8662\;\text{\AA}$) onwards, as detailed later.

The \texttt{muse\_standard} recipe performs a telluric correction on the MUSE cube by using a standard star. This was not done perfectly, and some residuals remain in the worst-affected telluric regions. We show the average MUSE spectra with the narrow telluric affected regions highlighted (grey shaded bands) in Figure \ref{fig:icspec}. These regions do not overlap with any features we used for analysis (described in Sections \ref{sec:moskin} and \ref{sec:icstellpops}).

Due to the large size of the data cube, the outer regions that were worst affected by sky and had only very low-S/N galaxy data were cropped. This left the central $30\arcsec \times 30\arcsec$, $150\times 150$~pixels centred on the galaxy core\footnote{The subtraction of the pseudo sky and cube cropping was done by Joshua Warren and is described in his DPhil thesis.}.

\section{Data Analysis}
\label{sec:analysis}

\subsection{Mosaic Kinematics}
\label{sec:moskin}

\subsubsection{Voronoi Binning}
\label{sec:voronoi}

To understand the three-dimensional structure of the galaxy, we used the spectra in the MUSE and KMOS data cubes to measure the stellar kinematics across the surface of the galaxy. As discussed, the data were taken in variable weather conditions for the four mosaics and as a poor-seeing backup target. Therefore, the quality of the spectra varied across the different mosaics. In order to preserve the data quality where possible, we adopted the Voronoi binning method of \cite{Cappellari2003} that adaptively bins the data into regions of a constant signal-to-noise ratio (S/N). The \textsc{voronoi\_2d\_binning}\footnote{\href{http://www-astro.physics.ox.ac.uk/\~mxc/software/}{\url{http://www-astro.physics.ox.ac.uk/~mxc/software/}}} software required a signal and noise for each spatial pixel across the mosaic. For the MUSE mosaic, this was done by taking the median value of the spectrum from each spatial pixel in the data cube and noise cube. 

For the KMOS cubes, identifying signal and noise values was more complicated as the noise cubes were not properly propagated by the pipeline. To estimate a signal and noise value for the KMOS mosaics, we fitted the continuum region of the CaT \citep[as defined by][]{Cenarro2001} with the strong sky lines masked. We took the average continuum level as our signal value and the standard deviation of the data to the fit as a measure of the noise for each spatial pixel of the KMOS mosaics. The mosaics were binned to different S/N values (between 20 and 100). Normalising and taking the median of all the spectra within these different S/N Voronoi bins gave us spectra to analyse at a common S/N value across all the mosaics from KMOS and MUSE. 

\subsubsection{Deriving the Kinematics}
\label{sec:derivkin}

To measure the stellar kinematics, we used the Penalised PiXel-Fitting method \citep[\textsc{ppxf};][]{Cappellari2004, Cappellari2017}. The templates we used to fit the spectra were the Extended Medium-resolution Isaac Newton Telescope Library of Empirical Spectra \citep[E-MILES;][]{Sanchez-Blazquez2006, Falcon-Barroso2011}-based simple stellar population (SSP) models of \cite{Vazdekis2016}\footnote{Available from \href{http://miles.iac.es/}{\url{http://miles.iac.es/}}.}. These have a broad spectral range ($\sim1680\text{--}50,000\;\text{\AA}$) making them ideal for both the optical MUSE and NIR KMOS data. The E-MILES library (and the SSP models based upon them) has a spectral resolution of $2.54\;\text{\AA}$ FWHM \citep[][]{Beifiori2011} over the range $3540\text{--}8950\;\text{\AA}$ increasing to $\sigma=60$~km~s$^{-1}$ at greater wavelengths. We used the E-MILES-based SSP models based on the Bag of Stellar Tracks and Isochrones \citep[BaSTI;][]{Pietrinferni2004} models for deriving kinematics due their high resolution.

\begin{figure}
\centering
\includegraphics[width=0.46\textwidth, trim=0 0 0 0 ,clip]{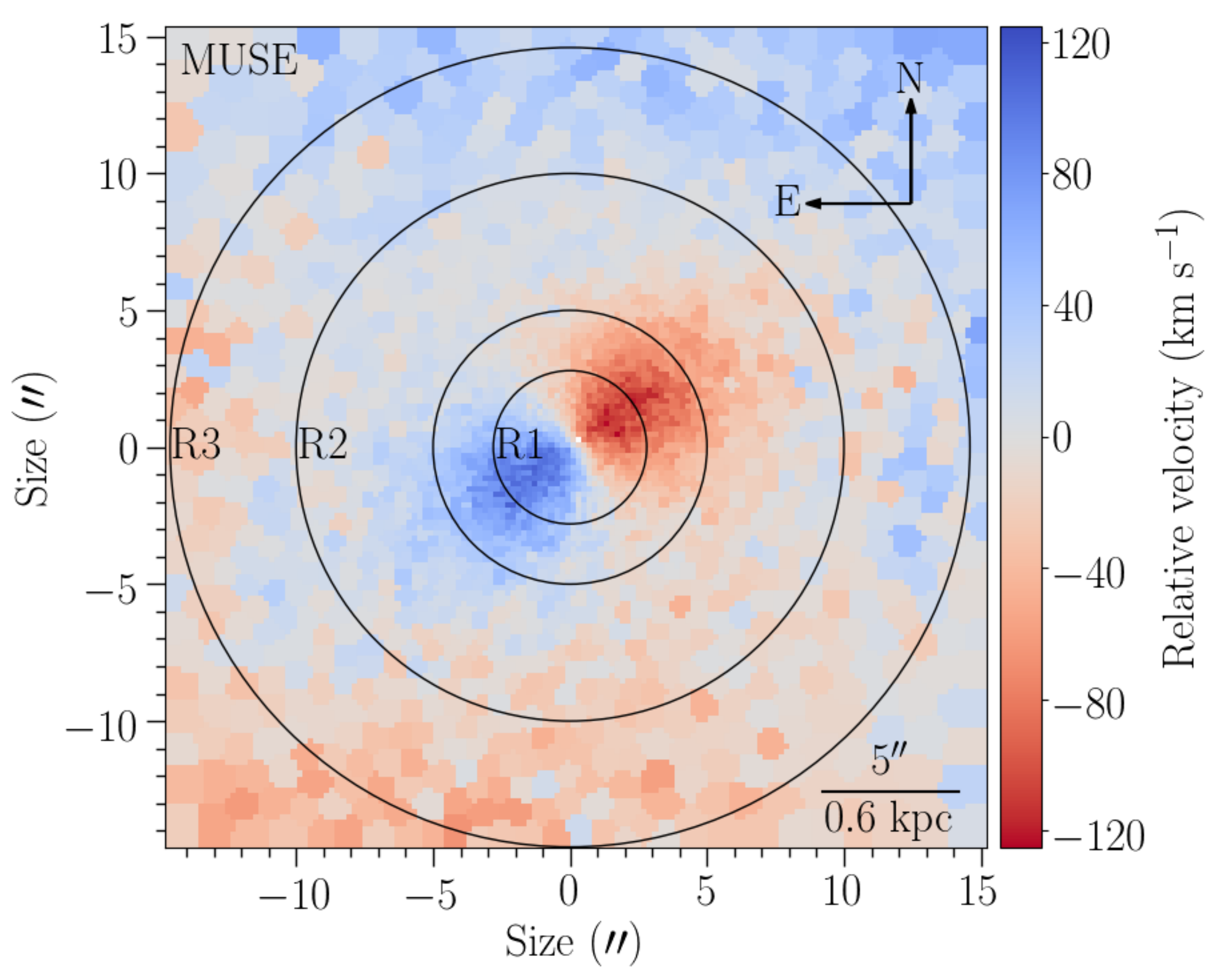}\vspace{1mm}
\includegraphics[width=0.46\textwidth, trim=0 0 0 0 ,clip]{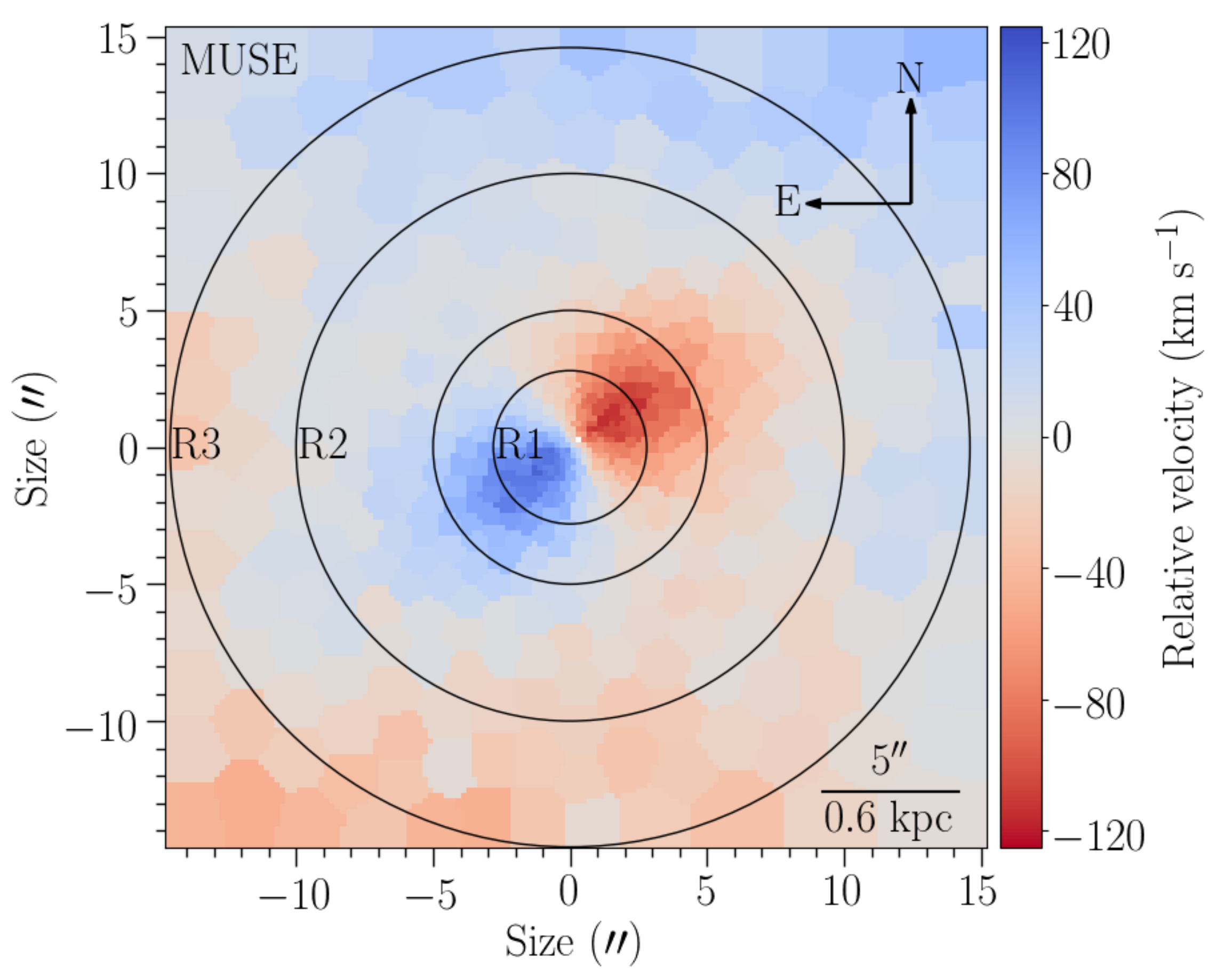}\vspace{1mm}\\
\includegraphics[width=0.46\textwidth, trim=0 0 0 0 ,clip]{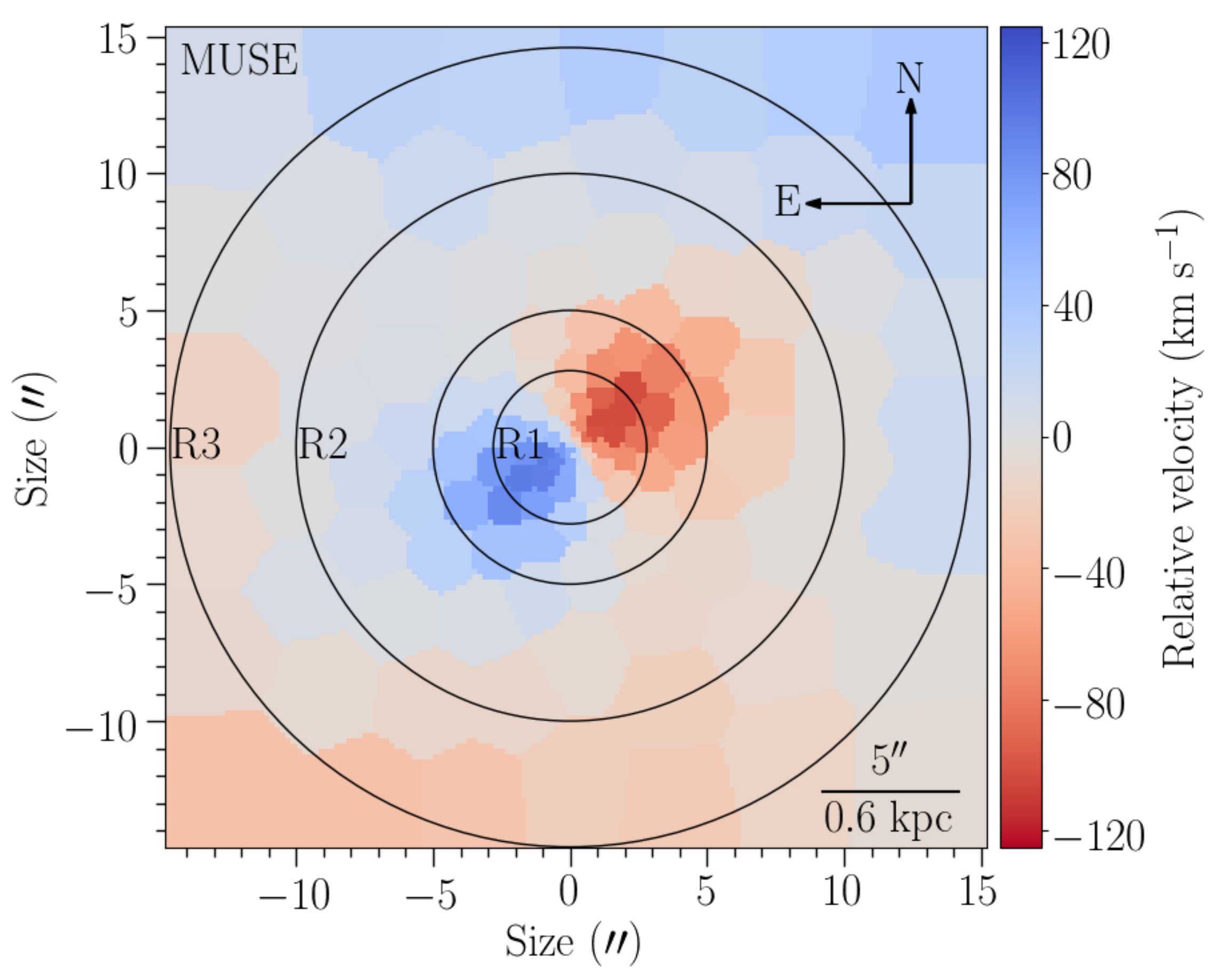}
\caption[MUSE relative velocity maps of IC~1459]{Relative velocity maps of IC~1459 as measured using \textsc{ppxf} on the MUSE cube. The mosaics are $150\times 150$ pixels and $30\arcsec \times 30\arcsec$ in size. The circles represent regions (R) from which spectra were extracted (Section \ref{sec:exspec}) and properties of the stellar populations derived (see Section \ref{sec:icstellpops}). Pixels in the cube were Voronoi binned to different S/N thresholds with the \textsc{voronoi\_2d\_binning} code \citep{Cappellari2003}. Spectra within the bins were normalised, the median taken and their kinematics fitted. \textit{Top:} Pixels binned to S/N~=~20.  \textit{Middle:} Pixels binned to S/N~=~40. \textit{Bottom:} Pixels binned to S/N~=~80. See Sections \ref{sec:moskin} and \ref{sec:exspec}.}
\label{fig:musekinbin}
\end{figure}

\begin{figure}
\centering
\includegraphics[width=0.45\textwidth, trim=0 0 0 0 ,clip]{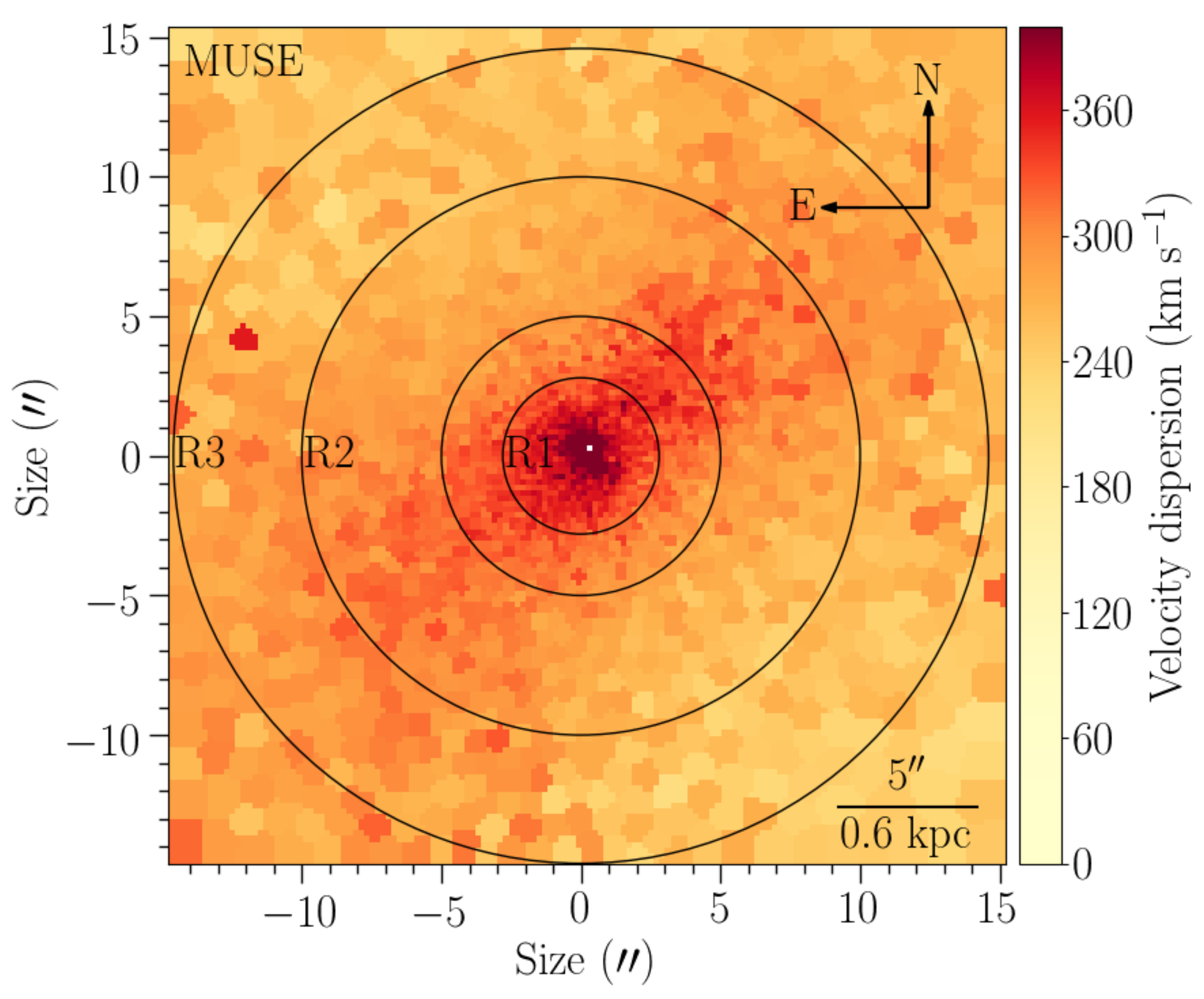}\vspace{1mm}
\includegraphics[width=0.45\textwidth, trim=0 0 0 0 ,clip]{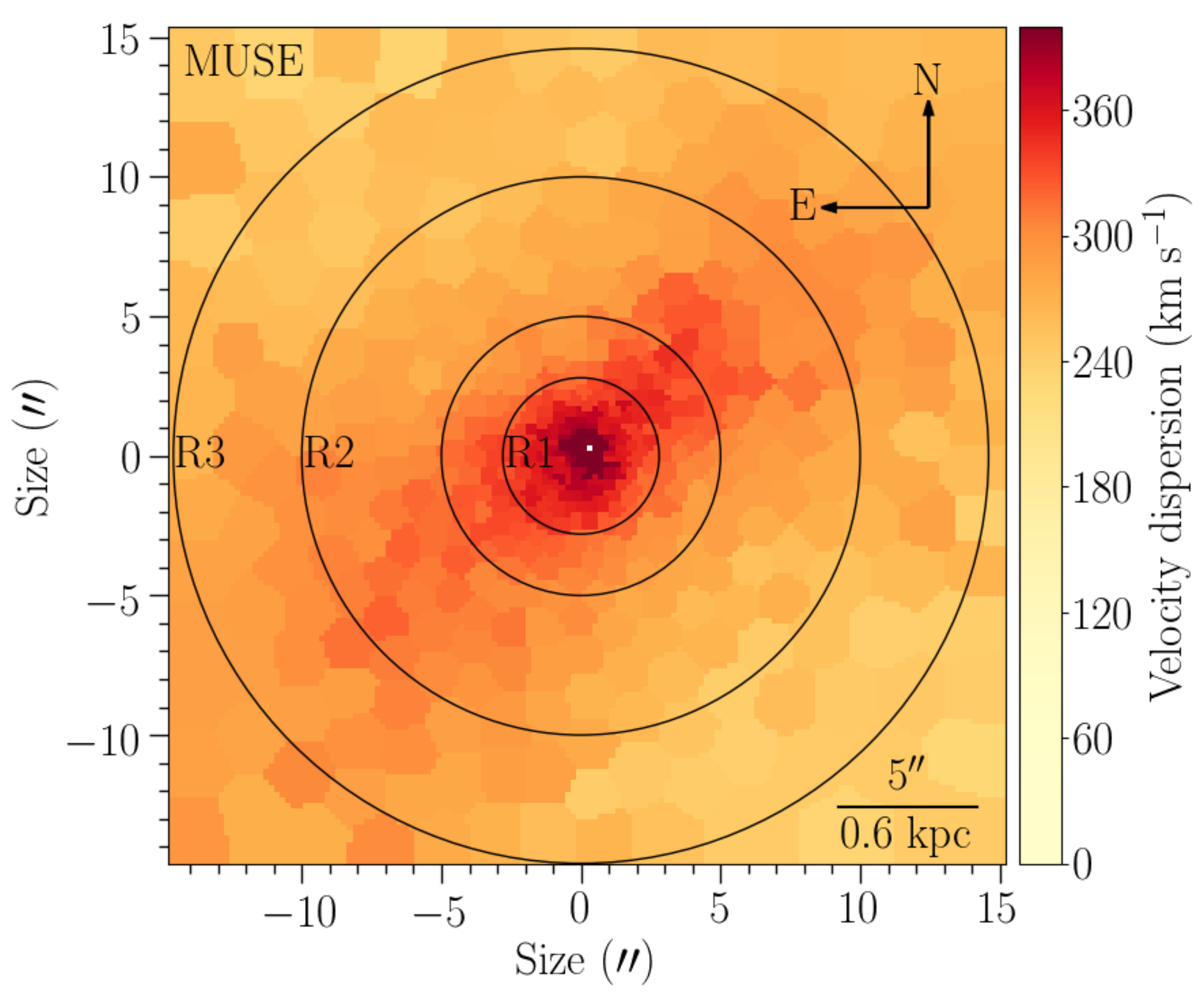}\vspace{1mm}
\includegraphics[width=0.45\textwidth, trim=0 0 0 0 ,clip]{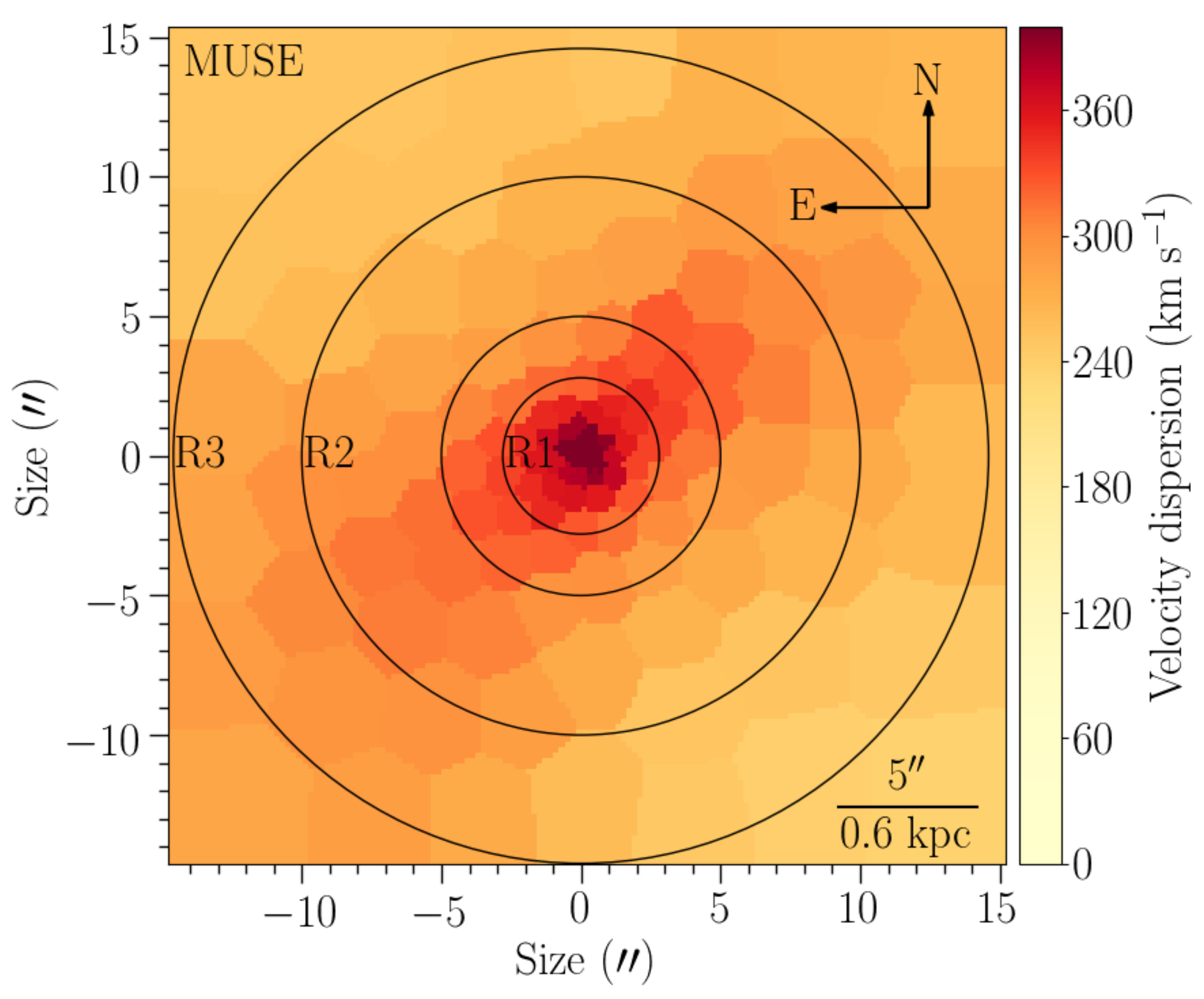}
\caption[MUSE velocity dispersion maps of IC~1459]{Velocity dispersion maps of IC~1459 as measured using \textsc{ppxf} on the MUSE cube with the same scales and labelling as Figure \ref{fig:musekinbin}. \textit{Top:} Pixels binned to S/N~=~20. \textit{Middle:} Pixels binned to S/N~=~40. \textit{Bottom:} Pixels binned to S/N~=~80. See Sections \ref{sec:moskin} and \ref{sec:exspec}.}
\label{fig:musesigbin}
\end{figure}

We fitted the spectra with a fourth-order additive polynomial (advised for extracting kinematics with \textsc{ppxf}) to the continuum. The MUSE spectra were fitted over its cleaner blue region with strong stellar-absorption features, between 4750--5850 $\text{\AA}$. We masked a single bright sky feature in this wavelength range (over 5570--5587 $\text{\AA}$). For the KMOS spectra we fitted nearly the whole spectrum between 8100--10400 $\text{\AA}$.

\begin{figure}
\centering
\includegraphics[width=0.46\textwidth, trim=0 0 0 0 ,clip]{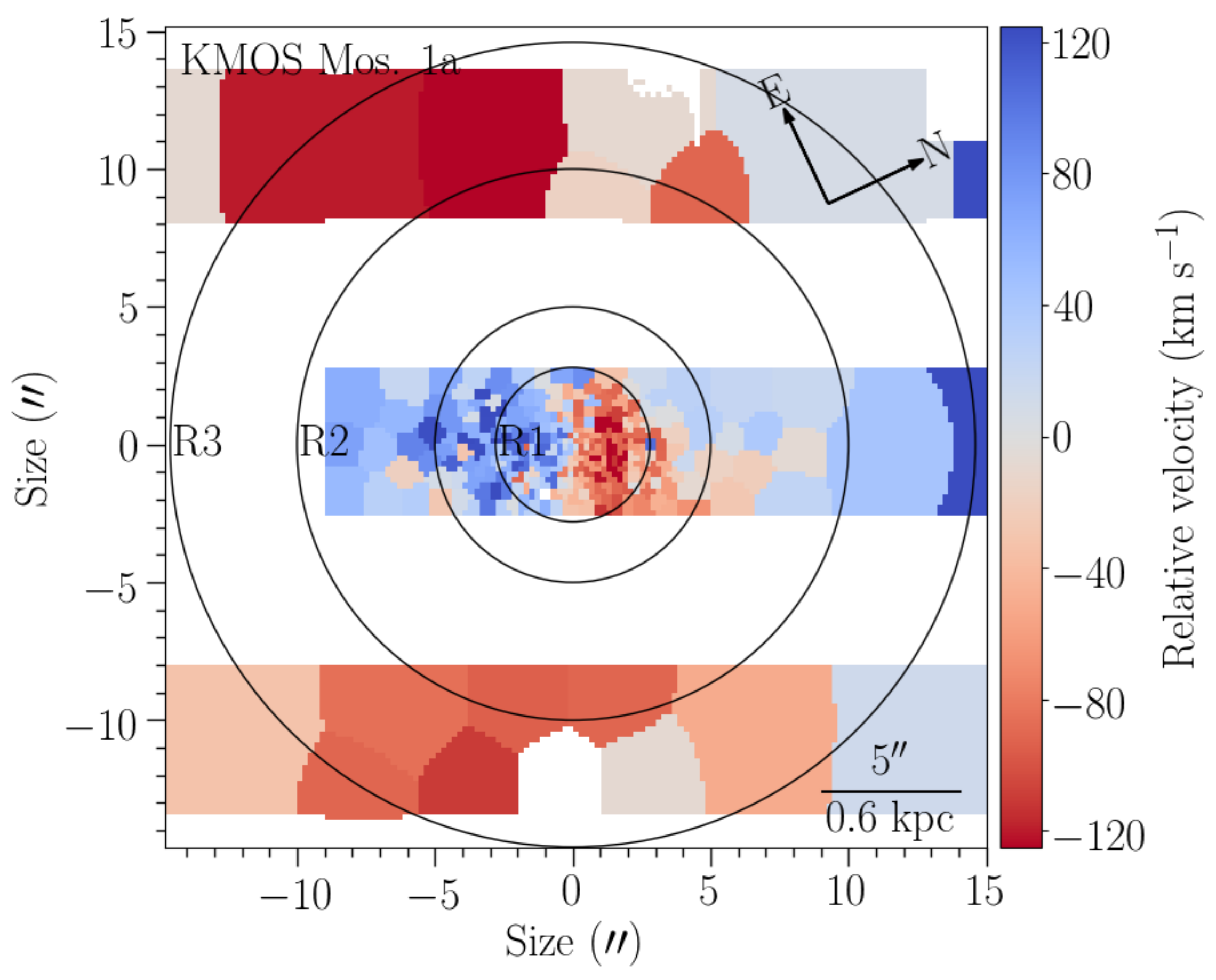}\vspace{1mm}
\includegraphics[width=0.46\textwidth, trim=0 0 0 0 ,clip]{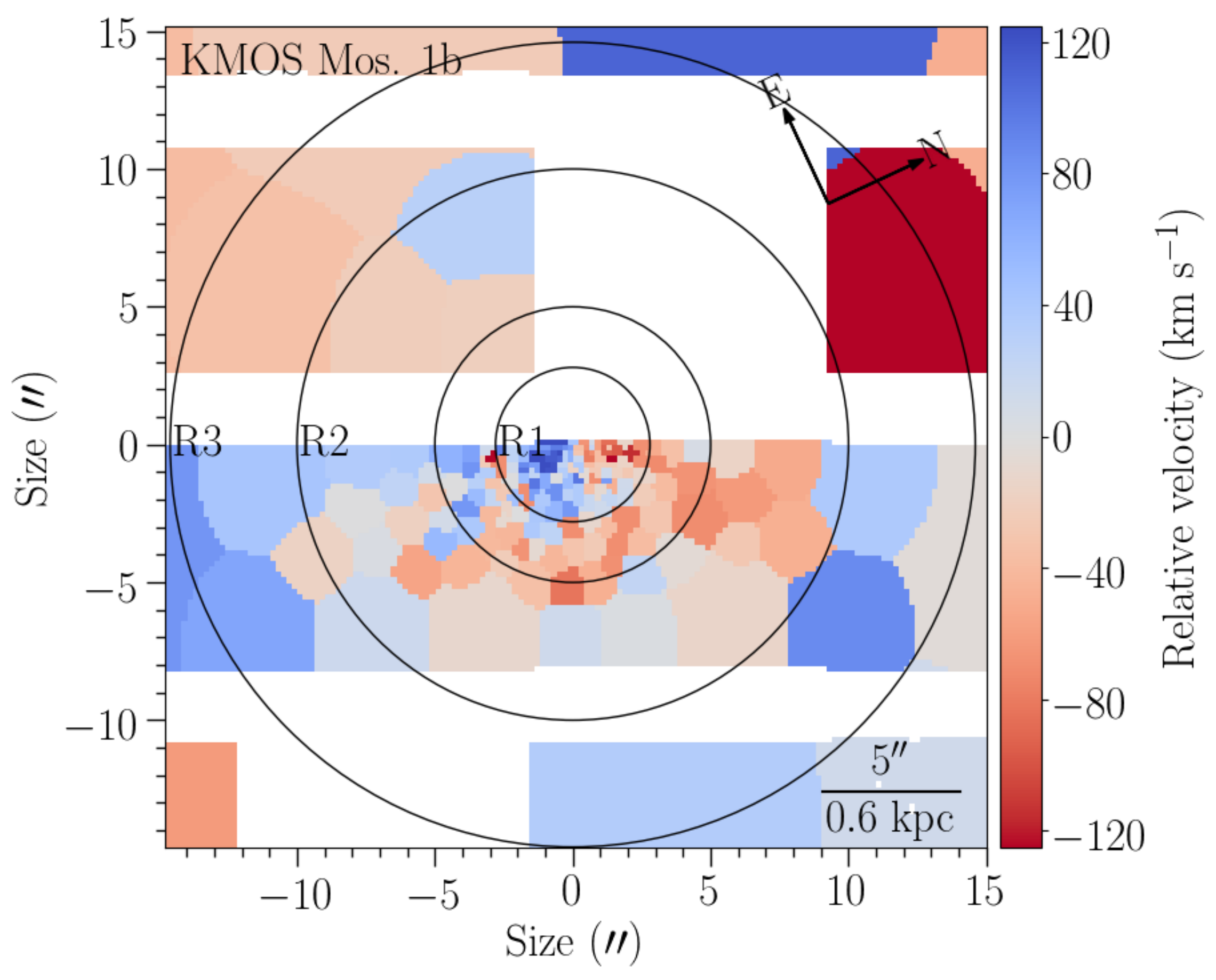}\vspace{1mm}
\includegraphics[width=0.46\textwidth, trim=0 0 0 0 ,clip]{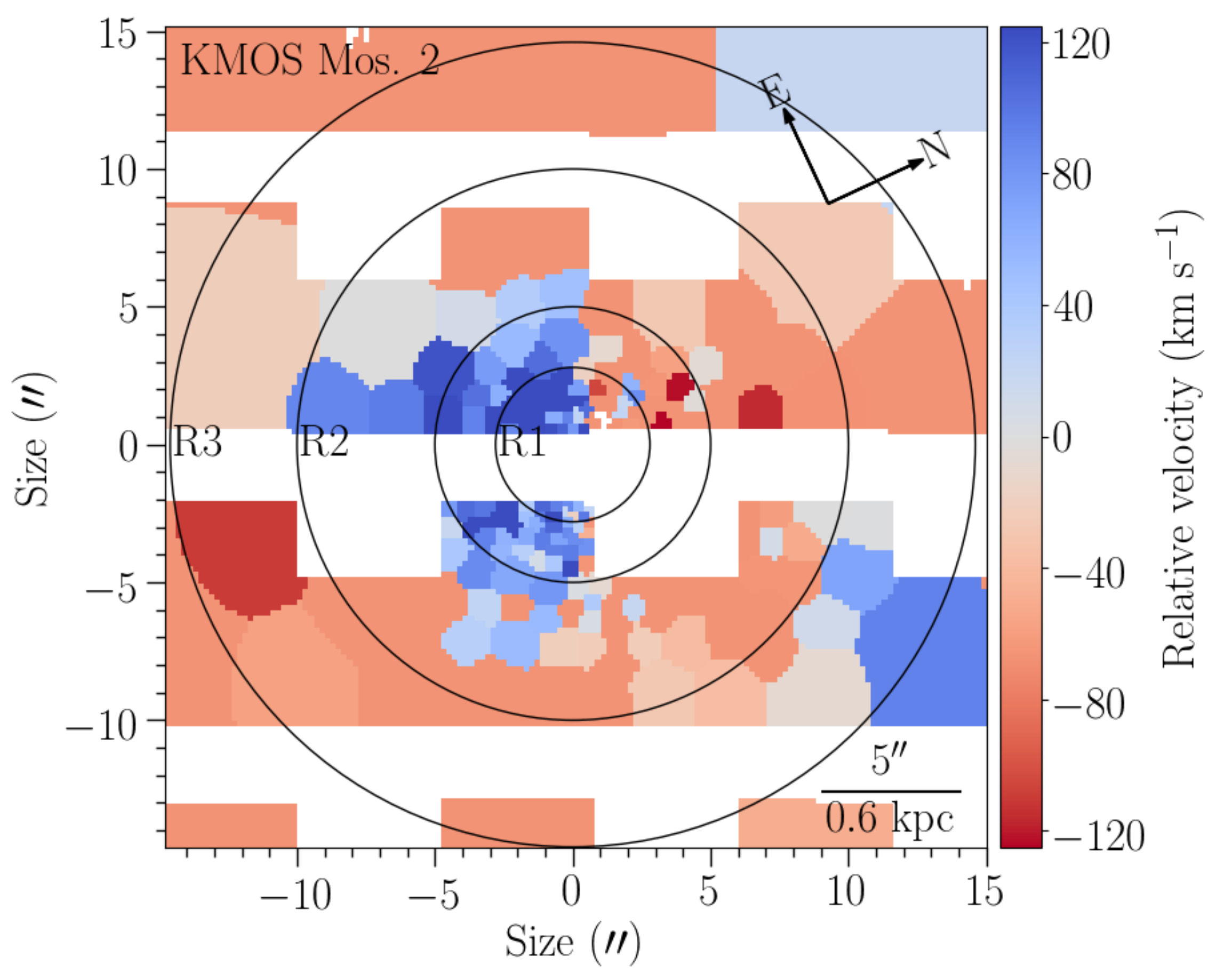}
\caption[KMOS relative velocity maps of IC~1459]{Relative velocity maps of IC~1459 as measured using \textsc{ppxf} on the KMOS mosaics. The circles represent the same regions of the galaxy as shown in Figure \ref{fig:musekinbin}. The white regions are NaN values where there is no data or a fit to the data was not possible. The KMOS mosaics were cropped to the size of the clipped MUSE cube. All mosaics were Voronoi binned to S/N~=~40. \textit{Top:} Mosaic 1a. \textit{Middle:} Mosaic 1b. \textit{Bottom:} Mosaic 2. See Sections \ref{sec:moskin} and \ref{sec:exspec}.}
\label{fig:kmoskinbin}
\end{figure}

\begin{figure}
\centering
\includegraphics[width=0.45\textwidth, trim=0 0 0 0 ,clip]{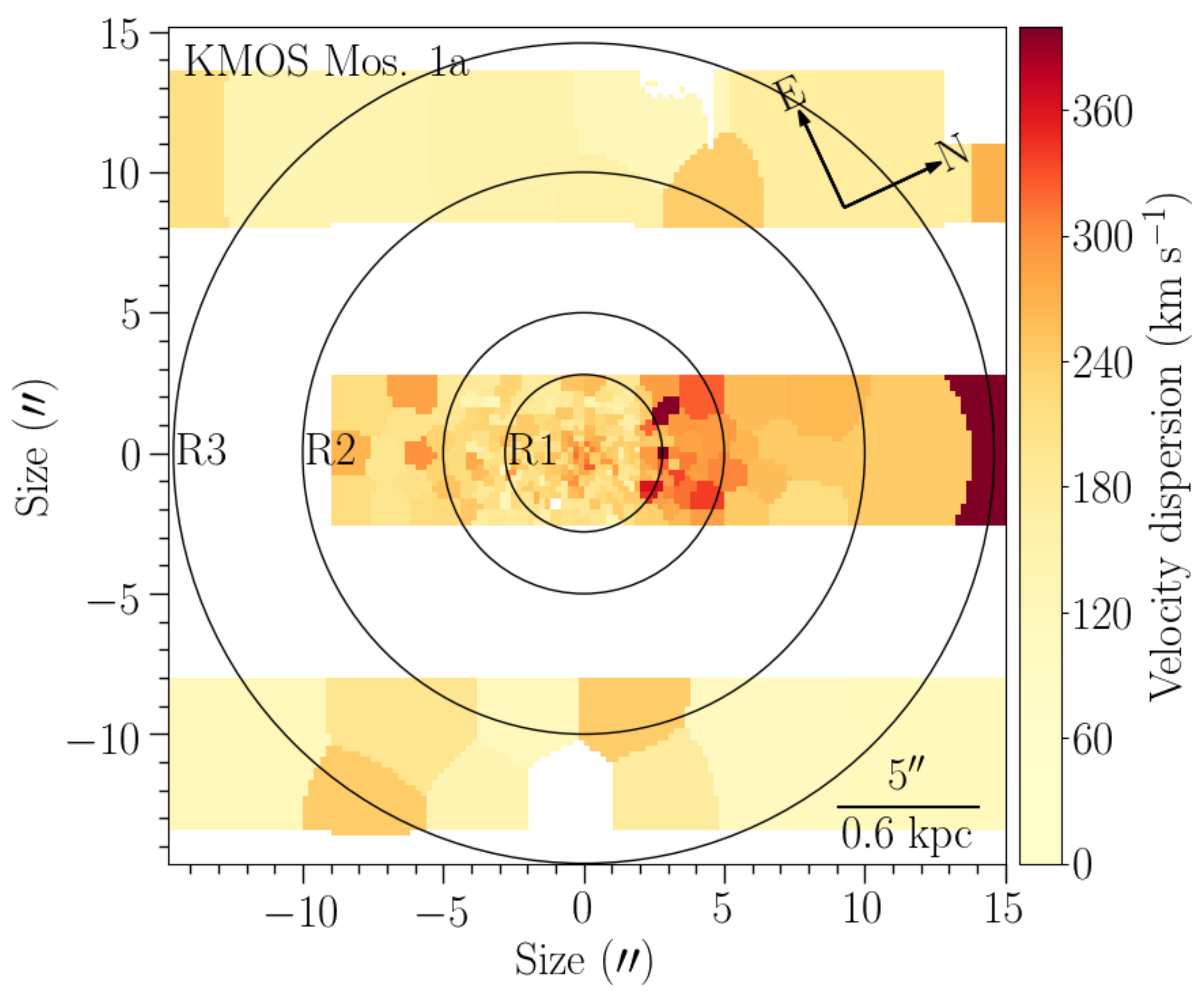}\vspace{1mm}
\includegraphics[width=0.45\textwidth, trim=0 0 0 0 ,clip]{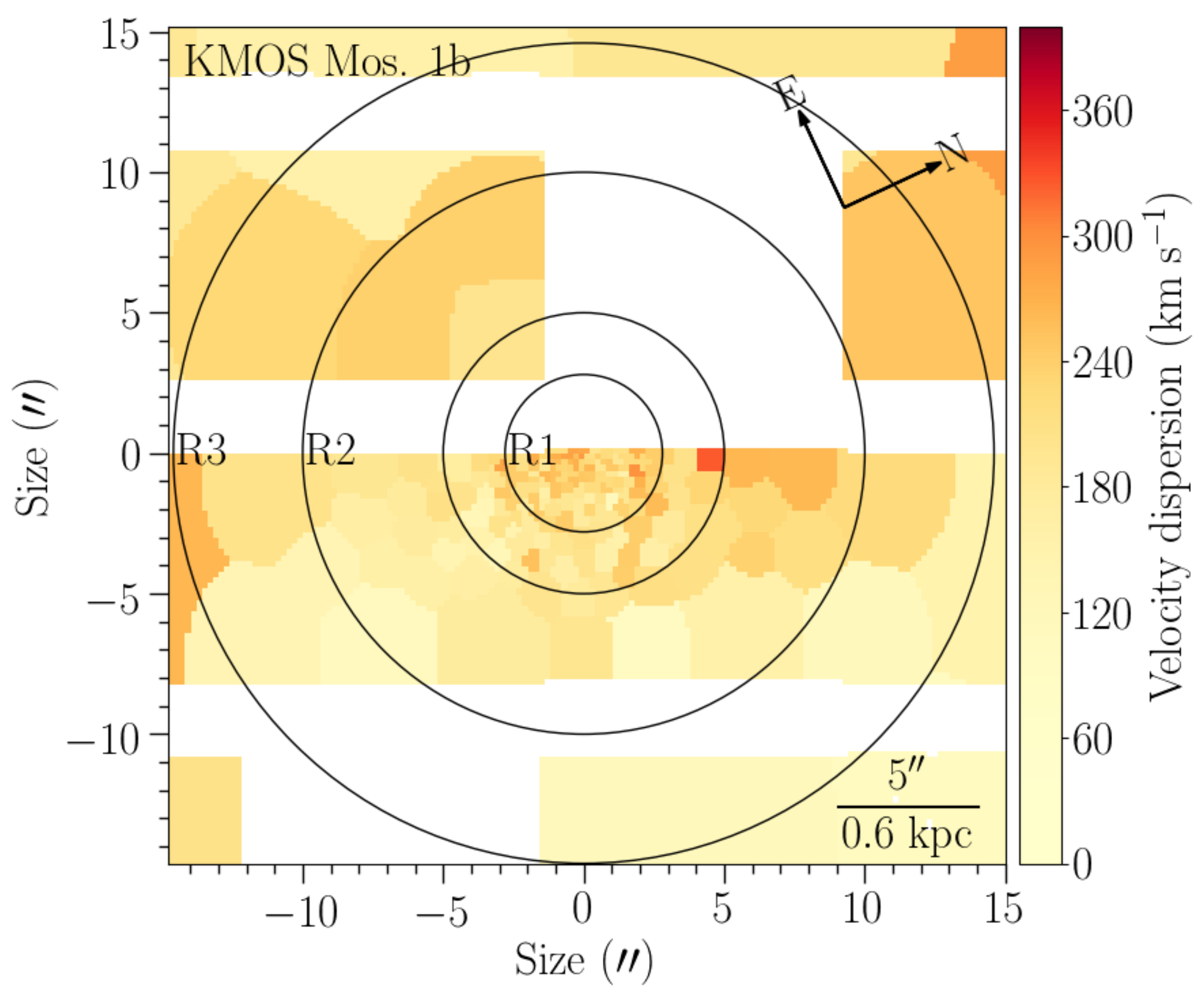}\vspace{1mm}
\includegraphics[width=0.45\textwidth, trim=0 0 0 0 ,clip]{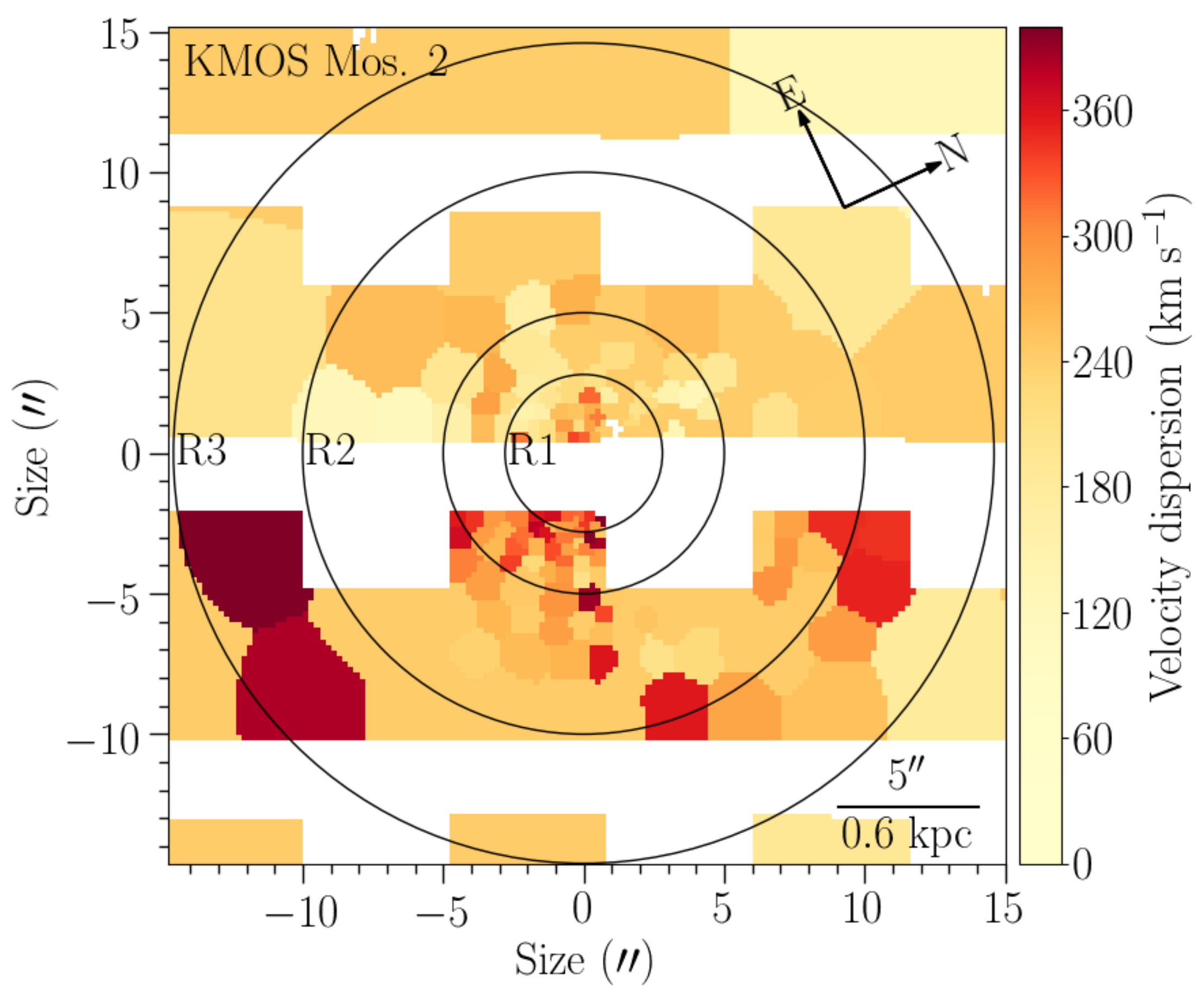}
\caption[KMOS velocity dispersion maps of IC~1459]{Velocity dispersion of IC~1459 as measured using \textsc{ppxf} on the KMOS mosaics with the same scale and labelling as Figure \ref{fig:kmoskinbin}. All mosaics were Voronoi binned to S/N~=~40. See Sections \ref{sec:moskin} and \ref{sec:exspec}.}
\label{fig:kmossigbin}
\end{figure}

\subsubsection{Kinematic Maps}
\label{sec:kinmap}

We show the velocity relative to the median velocity of IC~1459 ($\sim1700$~km~s$^{-1}$ as measured from MUSE) for MUSE and KMOS in Figures \ref{fig:musekinbin} and \ref{fig:kmoskinbin} respectively. The stellar velocity dispersion maps for MUSE and KMOS are shown in Figures \ref{fig:musesigbin} and \ref{fig:kmossigbin} respectively. We show the MUSE mosaic (full coverage over the central regions of the galaxy) with three Voronoi binning thresholds: S/N~=~20, 40, and 80 in Figure \ref{fig:musekinbin}. A striking feature of the relative velocity maps of MUSE is of course the strongly counter-rotating core in the central $\sim 10\arcsec$ ($R_{\rm KDC} \approx 5^{\prime\prime} \approx 0.1 R_{\rm e}$) and the slower rotation of the outer regions of the galaxy (seen starting at a radius of $\sim10\arcsec$). From the MUSE velocity maps, we measure peak rotation of the core of $\sim125$~km~s$^{-1}$ and $\sim-122$~km~s$^{-1}$, with errors around $\sim0.5\%$ as given in Table \ref{tab:errs} for MUSE. This is lower than previous estimates \citepalias{Franx1988}; the lower rotational velocity measured here is likely to be due to the poorer seeing. We show the kinematics of the three KMOS mosaics with S/N~=~40 Voronoi binning in Figure \ref{fig:kmoskinbin}. All the images are cropped to the same size as the MUSE mosaic, and the north-east arrows show the orientation of the galaxy in each.

The galaxy appears to have a higher average dispersion that extends from its centre to the south east. This may be a feature of the reduction, or a genuine kinematic feature of the galaxy. There also appears to be a peak or plateau in dispersion towards the north west of the centre. In both directions, there appears to be a flaring of the high dispersion outward from the centre, creating an hourglass shape along the diagonal. This is discussed further in Section \ref{sec:disckin}.

\begin{figure}
\centering
\includegraphics[width=0.5\textwidth]{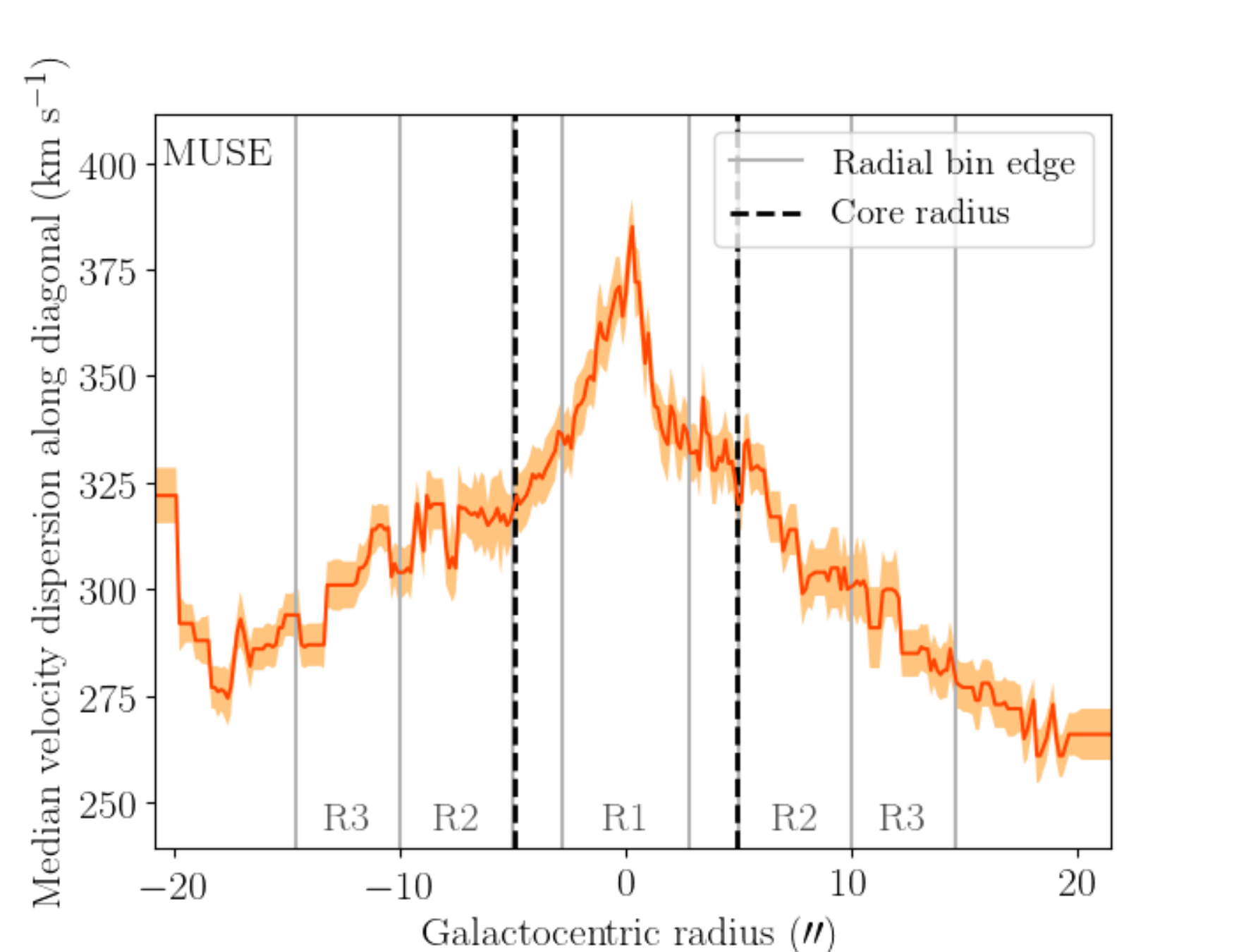}
\caption[]{Median velocity dispersion (dark-orange line) as a function of galactocentric radius for IC~1459. The median is measured along a $\sim5^{\prime\prime}$-wide diagonal slice (south-east to north-west corners) through the S/N $=20$ Voronoi-binned MUSE dispersion map (Figure \ref{fig:musesigbin}, top panel). The concentric radial bins on the kinematic maps are shown here (grey vertical lines, R1, R2, R3). The 1$\sigma$ errors (shaded orange region) and counter-rotating core radius (dashed vertical line) are shown.}
\label{fig:sigslice}
\end{figure}

To emphasise this point, we show the one-dimensional median dispersion measured from the S/N~$=20$ MUSE dispersion map, as a function of galactocentric radius in Figure \ref{fig:sigslice}. The median is taken along a $\sim5^{\prime\prime}$-wide diagonal slice (south-east to north-west corners) through the galaxy. The dispersion on the left of this plot (i.e. dominated by the asymmetry to the south east) is higher than that of the right (i.e. north west) outside of the core (black-dashed line). There is a shoulder in the dispersion to the right of the central peak coincident with the core radius. This could represent an unresolved secondary dispersion peak to the north west of the centre. This is seen most clearly in the S/N~$=80$ binned MUSE dispersion map (bottom plot of Figure \ref{fig:musesigbin}).

We determined errors for each Voronoi bin of the kinematic maps by adding random Gaussian noise to each wavelength pixel that scaled with its corresponding noise spectrum value. This new spectrum with added noise was then fitted with \textsc{ppxf} as before. This process was repeated 100 times so that 100 bootstrapped realisations of each spectrum were fitted with \textsc{ppxf}. From the distribution of output values, 1$\sigma$ errors were determined for each Voronoi bin. We quote the median of the percentage errors of each map in Table \ref{tab:errs}.

\begin{table*}
\centering
\caption{Median of the percentage error values of the kinematic maps (velocity -- $\delta v$, dispersion -- $\delta \sigma$) for IC~1459. The errors for the Voronoi bins of every map were determined from fitting 100 bootstrapped realisations of each spectrum and measuring a $1\sigma$ error from their distribution.}
\label{tab:errs}
\begin{tabular}{cccccccc}
\hline
\hline
 Mosaic & S/N & Median $\delta v$ & Median $\delta \sigma$ \\ \hline
MUSE & 20 & 0.5$\%$ & 3.1$\%$ \\
MUSE & 40 & 0.4$\%$ & 2.4$\%$ \\
MUSE & 80 & 0.3$\%$ & 2.0$\%$ \\
KMOS 1a & 40 & 1.7$\%$ & 12.8$\%$ \\ 
KMOS 1b & 40 & 1.6$\%$ & 12.9$\%$ \\ 
KMOS 2 & 40 & 2.3$\%$ &  16.6$\%$ \\ \hline
\end{tabular}
\end{table*}

The MUSE optical data and KMOS NIR data show similar stellar kinematics where it is possible to compare them. KMOS Mosaic 1a (top panels of Figure \ref{fig:kmoskinbin} and \ref{fig:kmossigbin}) is the only one with full coverage of the core and the counter rotation is seen clearly. There is a strong central dispersion peak seen in MUSE, but less so in KMOS, most likely due to the larger average errors for KMOS (Table 1). These larger relative errors and patchy coverage of the KMOS maps means that the dispersion asymmetry seen in MUSE is not identifiable in KMOS.

\subsection{Extracting Average Spectra}
\label{sec:exspec}

In all of the mosaics from MUSE and KMOS in Figures \ref{fig:musekinbin}, \ref{fig:musesigbin}, \ref{fig:kmoskinbin} and \ref{fig:kmossigbin}, concentric circles are shown of the same scale and approximate position relative to the galaxy. These circles were determined using the MUSE relative-velocity maps to encompass different regions of the galaxy from which we extracted spectra. We took all the spectra within each radial bin from each of the unbinned mosaics, MUSE and KMOS 1a, 1b, and 2. We normalised the spectra in each bin before taking a median spectrum. For the MUSE mosaic and the KMOS Mosaics 1a and 1b, the regions were centred on the brightest pixel. For KMOS Mosaic 2, as the centre of the galaxy was not covered by the data, the centre was approximated using the distribution of light in the median images of the cubes and comparison with the other kinematic maps. Given the relatively high seeing ($\sim1\text{--}2\arcsec$) of all the observations, if this positioning was off by up to $\sim \nicefrac{2}{3}$ one KMOS IFU then it should not make a difference to the output spectra. We therefore chose to include the spectra from Mosaic 2 to improve the coverage of the KMOS data.

\begin{figure*}
\centering
\includegraphics[width=\textwidth, trim=0 0 0 0,clip]{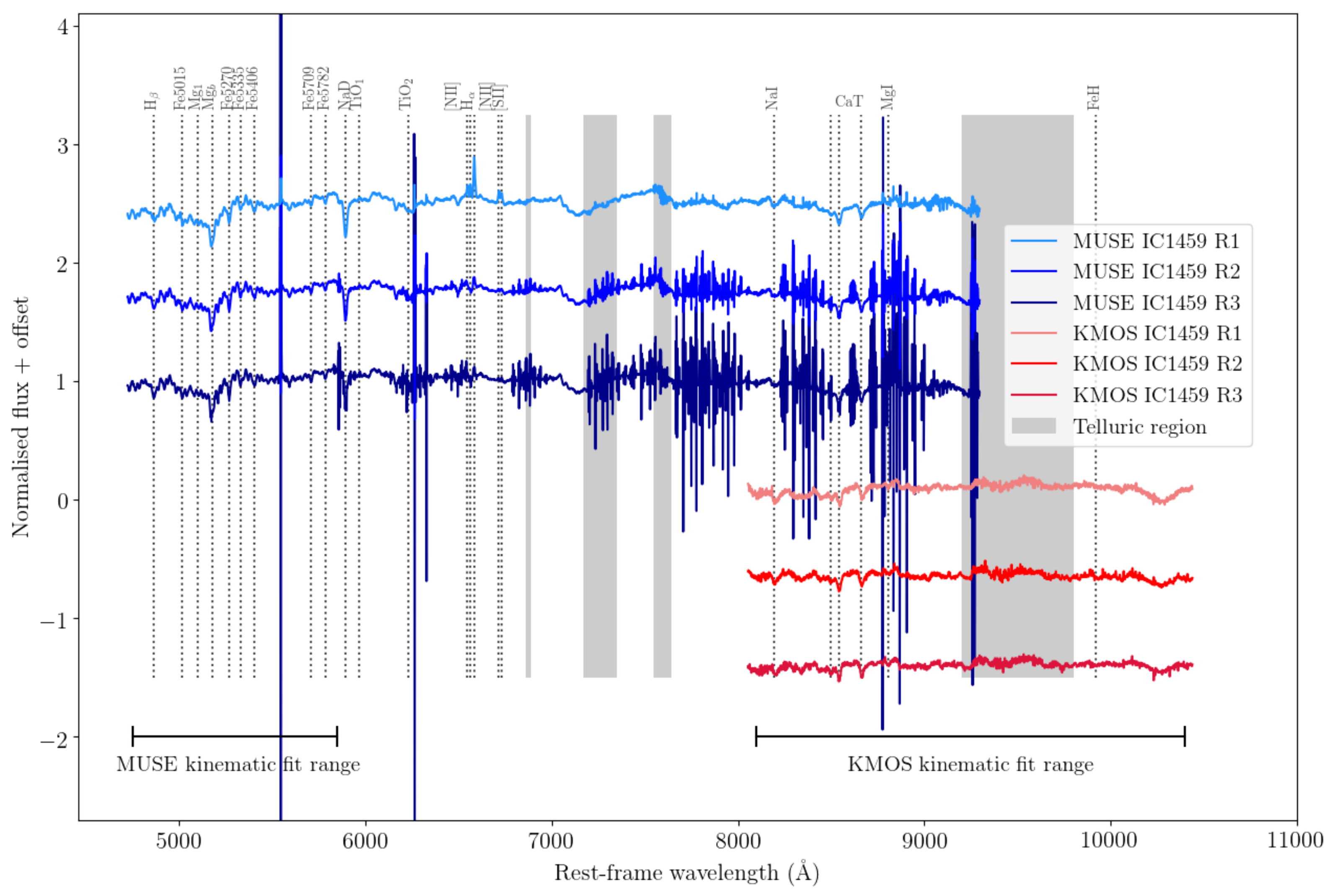}\\
\caption[Median MUSE and KMOS spectra of three radial bins of IC~1459]{Median spectra extracted from the three regions of IC~1459 indicated in Figures \ref{fig:musekinbin}, \ref{fig:musesigbin}, \ref{fig:kmoskinbin}, and \ref{fig:kmossigbin}. The MUSE spectra (blue) and KMOS spectra (combined from all three KMOS mosaics; red) are shown for the central region (R1; lightest shade), middle region (R2), and outer-most region (R3; darkest shade) of IC~1459. We show the spectral features of the galaxy (dotted lines) and regions of strong telluric absorption (shaded grey; see Section \ref{sec:tellmos}). All the strong emission not labelled is that of contamination from bright sky lines, this is strongest in the outer regions (R3) of the MUSE data. See Section \ref{sec:exspec}.}
\label{fig:icspec}
\end{figure*}

Using these uniformly defined regions across all the mosaics, we extracted a median spectrum from each by first normalising all the spectra within them. This left us with three spectra from the MUSE cube of regions R1, R2, and R3. For the KMOS data we had three sets of spectra for each region from the three mosaics. We further normalised these and made a single median KMOS spectrum for each of the three regions. In Figure \ref{fig:icspec} we show the final median spectra for each of the three regions as extracted from the MUSE (blue) and KMOS (red) mosaics. We indicate spectral features of the galaxy with dotted vertical lines. We indicate regions of the strongest telluric residuals (grey shaded regions) in both the optical (as spanned by MUSE) and NIR (spanned by KMOS where this correction was more important). These regions do not overlap with any of the features we used for the analysis of IC~1459. We also indicate the fitting regions used for analysis of both KMOS and MUSE.

All strong spectral features not indicated are sky lines that have resulted from poor sky subtraction. The sky features are particularly prominent in the outer regions of the MUSE mosaic (see Section \ref{sec:museprep}) and get stronger at redder wavelengths. Most of the valuable features for extracting kinematics and measuring stellar populations are much less affected at the bluer end of the MUSE spectrum (see Section \ref{sec:icstellpops}). The KMOS spectra are less contaminated by sky emission as the data were taken during dark time unlike the MUSE observations. The multi-epoch KMOS observations also helped to reduce variable noise when the spectra were stacked. 

Combining a large number of spectra results in unrealistically high S/N values for the extracted spectra when in reality, the noise between pixels is correlated. To quantify an appropriate S/N value for these combined binned spectra, we reverted to determining the S/N from the continuum around the CaT feature for both KMOS and MUSE (as discussed in Section 3.1.1). For MUSE we combined the noise spectra from the error cube by normalising and taking the median, as for the galaxy spectra. We then scaled the noise to the galaxy spectra using the S/N value determined from the CaT region. For KMOS with no error cube, we took the error spectrum to be the spectrum itself scaled to the S/N value determined from the CaT.

\subsection{Spatially Resolved Stellar Populations and Initial Mass Function}
\label{sec:icstellpops}

\subsubsection{Full-Spectral Fitting with \textsc{PyStaff}}
\label{sec:fullspec}

Using the spectra extracted from the radial bins across the surface of IC~1459, we measured the stellar populations of its different kinematic components. Combining the spectral range of both MUSE and KMOS, we were able to measure the stellar populations in the optical and the IMF sensitive features in the NIR. To fit the spectra, we used the full-spectral fitting code \textsc{PyStaff} (Python Stellar Absorption Feature Fitting)\footnote{Available from: \href{https://github.com/samvaughan/PyStaff}{https://github.com/samvaughan/PyStaff}} described in \cite{Vaughan2018b}, which we will briefly summarise here. We fit four spectral regions concurrently; between 4700--5600\,\AA, 5600--6800\,\AA, 8100--9000\,\AA~and 9700--10100\,\AA. In each region, we fitted the continuum with multiplicative polynomials of order $(\lambda_{\mathrm{high}}-\lambda_{\mathrm{low}})/100$. 

The code utilises the state-of-the-art stellar population synthesis models from \cite{Conroy2018}. These templates span from optical to NIR wavelengths (0.37--2.4$\;\mu$m) making them ideal for fitting the KMOS and MUSE data simultaneously. Using empirical spectra from the Extended-IRTF \citep{Villaume2017} and MILES \citep{Sanchez-Blazquez2006} stellar libraries, the templates model a single stellar population with age between 1--13.5 Gyr, total metallicity ([Z/H] = [Fe/H] + $A$[$\alpha$/Fe], where the constant $A$ depends on the model abundance pattern; e.g. \citep{Conroy2013}) from -1.5 to +0.2 dex and a variable low-mass IMF. The IMF is modelled as a two-part power-law between 0.08 M$_{\odot}$ to 0.5 M$_{\odot}$ and 0.5 M$_{\odot}$ to 1.0 M$_{\odot}$, with each index a free parameter. The IMF above 1.0 M$_{\odot}$ is always fixed to a Salpeter slope ($x=2.3$).  The models also use synthetic `response functions' to account for variations in a number of individual elemental abundances. For further discussion of the templates and their construction, see \cite{Conroy2018}. 

The fit to the spectra of IC~1459 assumes a single stellar population as a star formation history (SFH). The model has 26 free parameters including galaxy recession velocity, velocity dispersion, stellar age, metallicity, and the slope of the two-part low-mass IMF. We also allow for variation in 10 common elemental abundances, most notably [Na/Fe] which is often enhanced with respect to the solar neighbourhood in massive ETGs. 

\begin{figure*}
\centering
\includegraphics[width=0.91\textwidth]{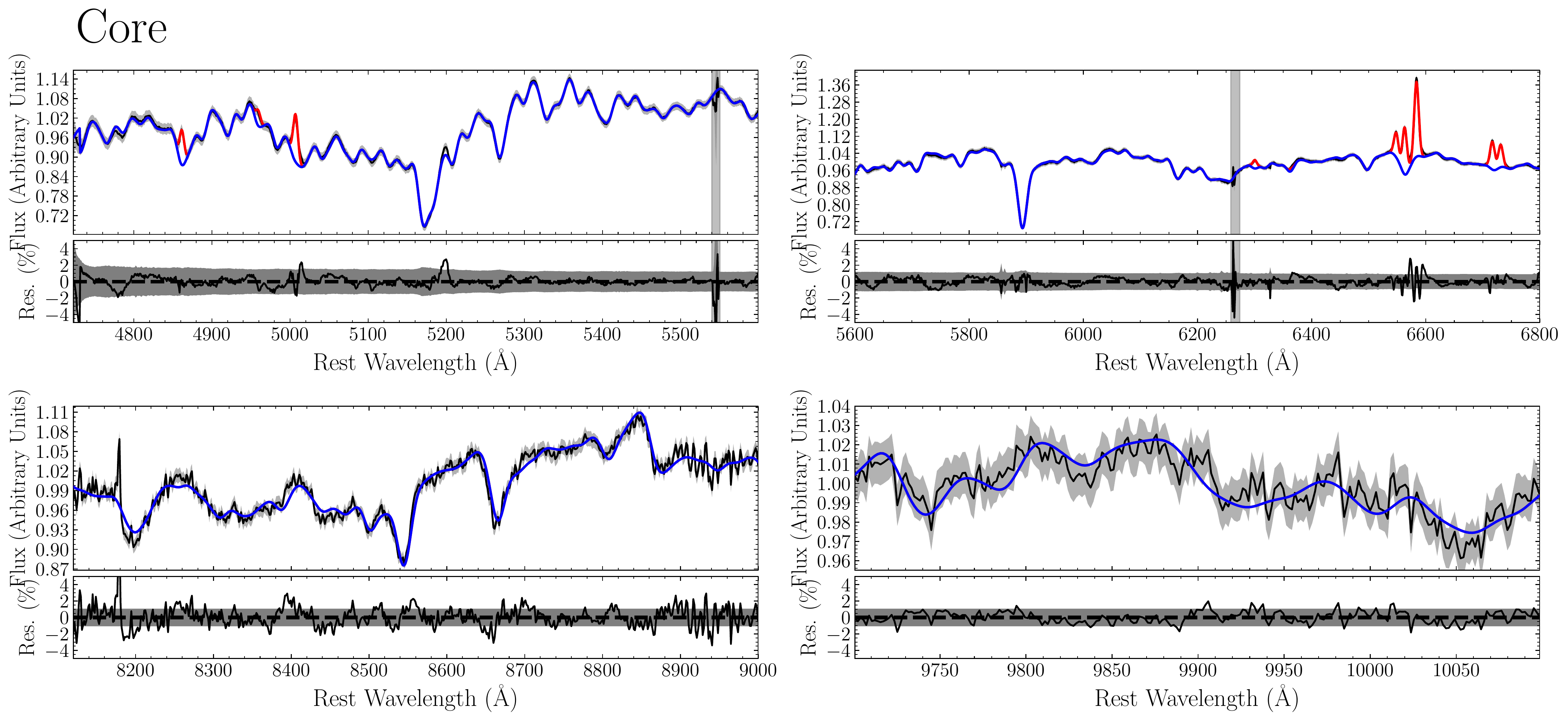}
\includegraphics[width=0.91\textwidth]{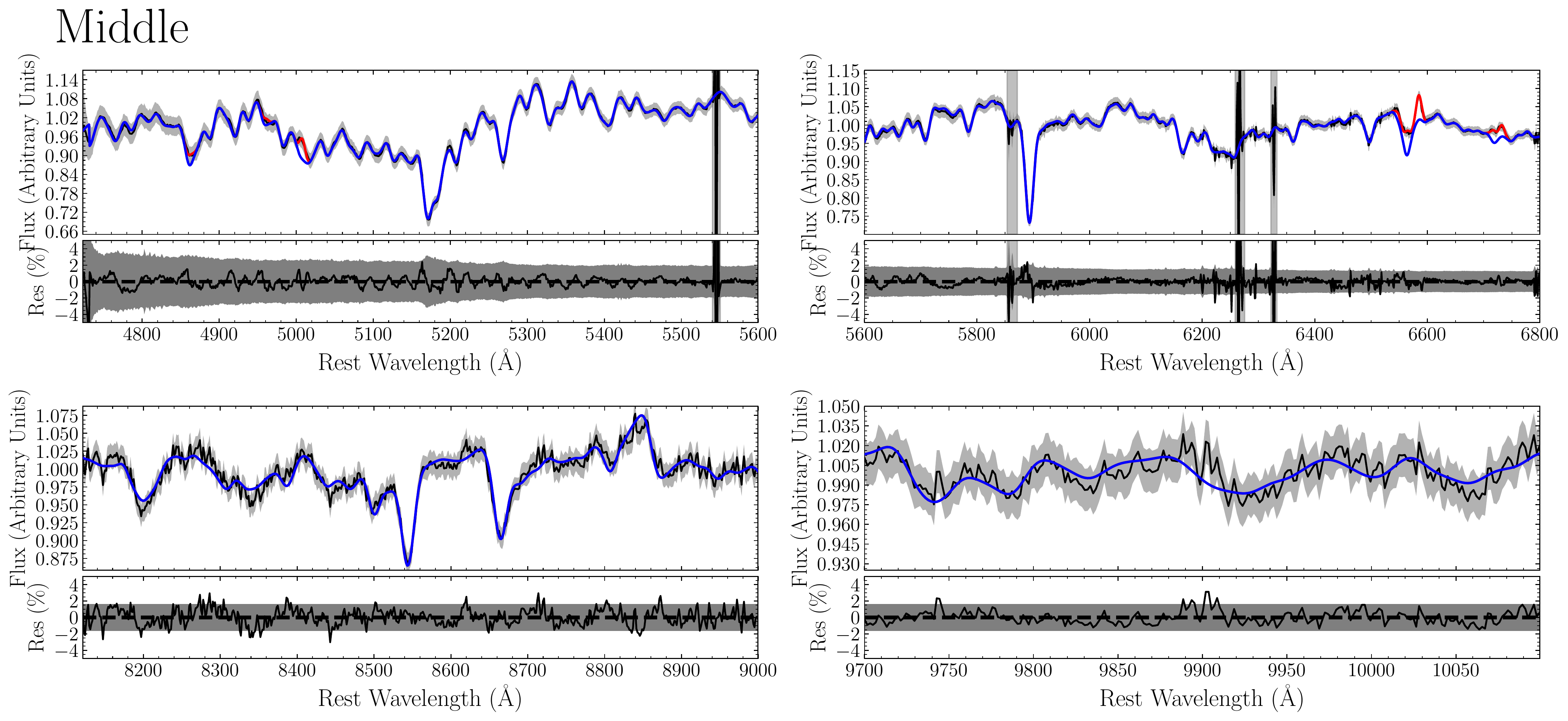}
\includegraphics[width=0.91\textwidth]{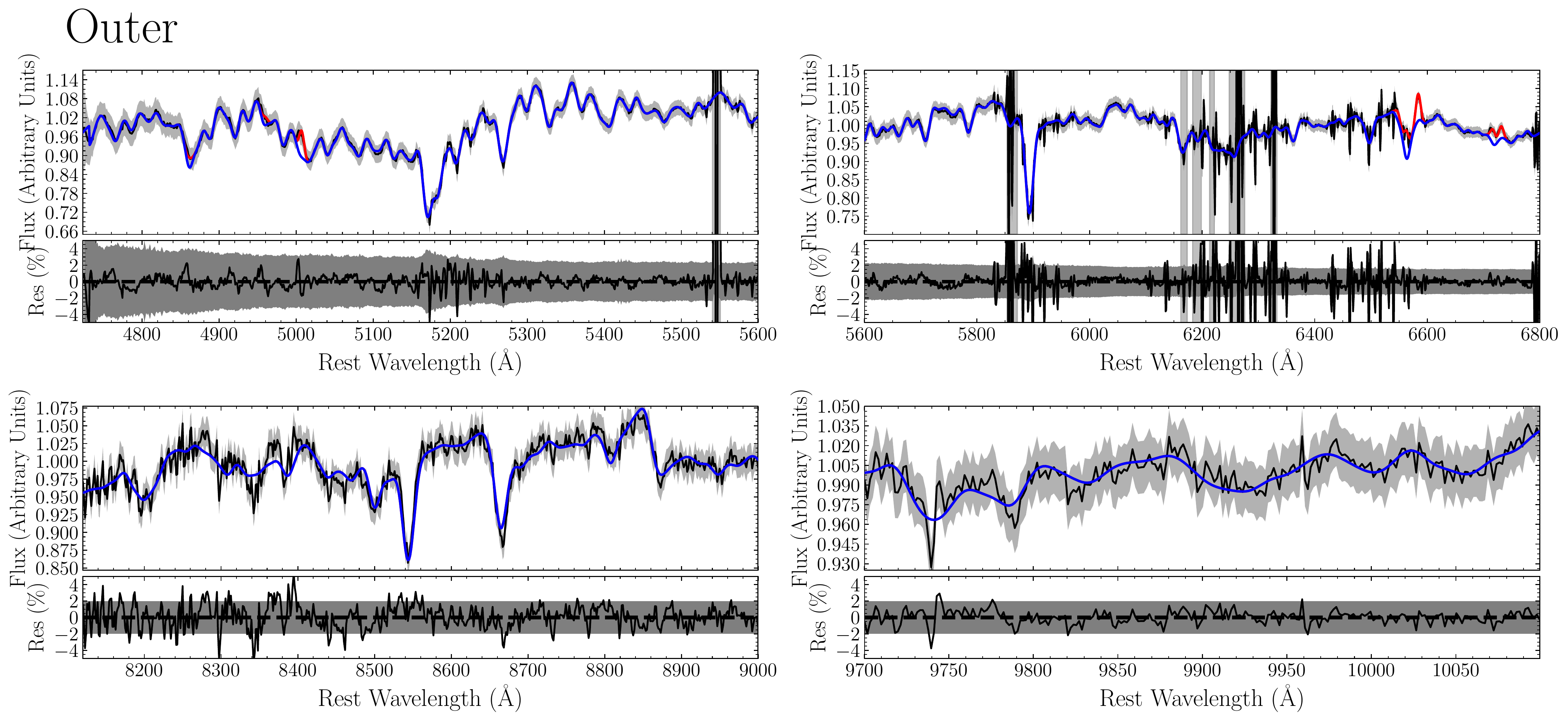}
\caption[]{Each set of four panels shows the \textsc{PyStaff} fit to the MUSE data (top two panels) and KMOS data (lower two panels) in four wavelength regions. The data (black), fit to the stellar component (blue) and fit to the gas emission component (red) are shown. The grey shaded regions represent the errors and the lower section of each panel shows the residuals of the fit to the data. Stellar population and IMF parameters and their uncertainties derived from the fits are summarised in Table \ref{tab:fsstellpop}. \textit{Top:} Fits to the core region (R1) of IC~1459. \textit{Middle:} Fits to the middle region of IC~1459 (R2). \textit{Bottom:} Fits to the outer region of IC~1459 (R3). See Section \ref{sec:fullspec}.}
\label{fig:fullspecfit}
\end{figure*}

Strong Balmer and forbidden emission lines can be seen in the optical wavelengths spanned by the MUSE data (see Figure \ref{fig:fullspecfit}). This is most apparent in the core of IC~1459 (R1) in the MUSE data as exhibited by the H$\alpha$, H$\beta$, [NII], [SII], line and [OIII]$\lambda5007\;\text{\AA}$ emission in Figure \ref{fig:icspec}. We therefore include the strength of seven common emission lines in the fit.  These emission lines are tied to have the same velocity and velocity dispersion, which can be different from the kinematic parameters of the stellar component.

The parameter space was explored using the Markov-chain Monte-Carlo (MCMC) code \texttt{emcee} \citep{Foreman-Mackey2013}. We ran each fit with 200 walkers for 100,000 steps, discarding the first 25,000 steps as the "burn-in", and visually inspected each chain for convergence. We also ensured that the posterior surface recovered from the fitting was smooth, using corner plots to ensure it showed no bimodalities or other irregularities. 

\begin{figure}
\centering
\includegraphics[width=0.49\textwidth, trim=0 10 0 0, clip]{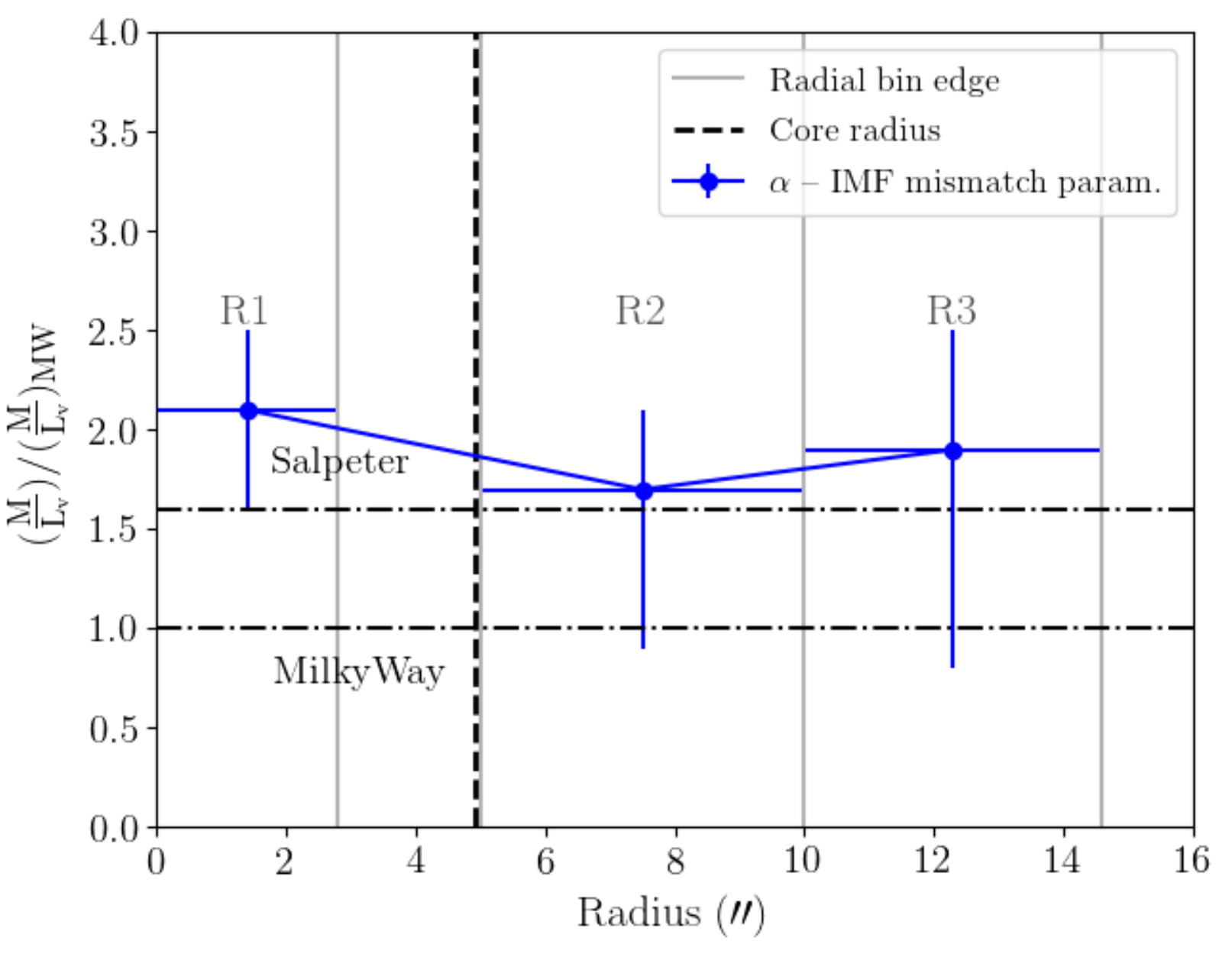}\vspace{5mm}
\caption[]{The IMF `mismatch' parameter $\alpha$  measured with \textsc{PyStaff} in the three radial bins: the galaxy core (R1), middle (R2) and outer (R3) regions (grey lines). The counter-rotating core boundary (dashed black line), and $\alpha$ values consistent with a Milky Way (MW)- and Salpeter-like IMF (dot-dash lines) are shown. The measurements show a roughly constant and bottom-heavy IMF ($\alpha \approx 2$) across all three bins, and the relatively large errors make them consistent with a Salpeter IMF ($\alpha \approx 1.6$). The limit of R3 probes to $\sim\nicefrac{1}{3}R_{\rm e}$, so this flat trend is consistent with similar studies \citep[e.g.,][]{vanDokkum2017, Vaughan2018b}. See Section \ref{sec:fullspec}.}
\label{fig:alpha}
\end{figure}

Figure \ref{fig:fullspecfit} shows the fits to the four wavelength regions. Each set of four panels represents a different radial bin of IC~1459; top four panels are the core (R1), middle four panels are from the middle (R2) and the bottom four panels are the outer regions (R3) of IC~1459. The top two panels in each set of four are the MUSE spectra the bottom two panels of each set of four are the KMOS spectra. The data (black), the fit to the stellar component (blue) and gas emission lines (red) are shown for each. The grey shaded regions represent the errors and the lower sections of each panel show the residuals of the data and fit. 

The values derived from the fits shown in Figure \ref{fig:fullspecfit} are summarised in Table \ref{tab:fsstellpop}. Results from \textsc{PyStaff} indicate a negative age gradient and relatively old ages ($\sim 9.4$--7.6~Gyrs). There is a slight negative metallicity gradient although it is also consistent with being constant within the errors.

We also measured the IMF `mismatch' parameter ($\alpha$). This is the ratio of the best-fitting mass-to-light ratio ($M/L$) and the $M/L$ for the galaxy with the best-fitting age and metallicity but with a Milky Way (MW)-like IMF. A galaxy with a MW-like IMF will therefore have an $\alpha$ value of $\sim 1$. Figure \ref{fig:alpha} shows the mismatch parameter value in the three regions of IC~1459. The IMF mismatch parameter measured from \textsc{PyStaff} is $\alpha\sim2.0$ to $\sim\nicefrac{1}{3}R_{\rm e}$, which is the limit of the outer radial bin (R3). The IMF is therefore more bottom heavy than a Salpeter IMF, although still consistent within the relatively large errors. The IMF for IC~1459 indicates it is significantly more dwarf-star dominated than the MW. These results are discussed in the context of the literature in Section \ref{sec:icdisc}.

The quality of the spectra in the fourth spectral region, around FeH, are of visibly lower quality. We ran a series of tests to verify what, if any, dependence including this region of the spectrum (9700--10100 $\text{\AA}$) had on our derived results. We ran \textsc{PyStaff} again for the MUSE and KMOS spectra for each radial bin without this fourth region. In the two outer bins (R2 and R3), the conclusions were unchanged. The resulting $\alpha$ value for each was consistent with the `fiducial' $\alpha$ value from the fit with all four wavelength regions, but with larger error bars. However, in the central bin (with the highest S/N), we found that discarding the FeH region did make a  difference to the inferred IMF. Without the FeH region, the best-fitting $\alpha$ value was $2.91^{+0.76}_{-0.65}$ whereas with the FeH region we found $\alpha = 2.06^{+0.49}_{-0.37}$. The age and metallicity values found were identical in both cases. It could be that the data in this region is of insufficient quality and is driving $\alpha$ down artificially in the centre of IC~1459, however, this is the region with the highest S/N. It is not clear why FeH should have an effect on $\alpha$ in the central radial bin alone (although the values remain consistent within the large errors). Previous studies have shown that FeH is difficult to model and is not well understood as a measure of the IMF which may be driving this difference \cite[e.g.][]{vanDokkum2017}.

\begin{table*}
\centering
\caption{Stellar population properties derived from the median MUSE and KMOS spectra in three regions of IC~1459 (left column) using \textsc{PyStaff} (middle three columns) and \textsc{ppxf} (last two columns). The age, total metallicity ([Z/H]) and IMF `mismatch' parameter ($\alpha$) derived from \textsc{PyStaff} are shown. We also show the mass-weighted age ($\langle\text{Age}\rangle$) and mass-weighted total metallicity ($\langle\text{[M/H]}\rangle$) derived from spectral fitting \textsc{ppxf} shown along with their errors. See Sections \ref{sec:fullspec} and \ref{sec:ppxf}.}
\label{tab:fsstellpop}
\begin{tabular}{cccccc}
\hline
\hline
Region & \multicolumn{3}{c}{\textsc{PyStaff}} & \multicolumn{2}{c}{\textsc{ppxf}}\\
 & Age (Gyrs) &  [Z/H] (dex) & $\alpha$ & $\langle\text{Age}\rangle$ (Gyrs) &  $\langle\text{[M/H]}\rangle$ (dex) \\\hline
R1 & $9.4 \substack{+0.8\\-0.7}$ & $0.156 \pm 0.017$ & $2.1 \substack{+0.5\\-0.4}$ & $12.0 \pm 0.5$ & $0.290 \pm 0.005$\\
R2 & $8.2 \substack{+1.1\\-1.0}$ & $0.150 \substack{+0.027\\-0.028}$ & $1.7 \substack{+0.8\\-0.4}$ & $11.3 \pm 0.4$ & $0.273 \pm 0.010$\\
R3 & $7.6 \substack{+1.3\\-1.1}$ & $0.142 \substack{+0.030\\-0.033}$ & $1.9 \substack{+1.1\\-0.6}$ & $11.1 \pm 0.6$ & $0.239 \pm 0.016$\\\hline
\end{tabular}
\end{table*}

\subsubsection{Fitting Stellar Populations with \textsc{ppxf}}
\label{sec:ppxf}

The main results from this work come from \textsc{PyStaff} due to the ability to vary the IMF while also fitting age and metallicity within a full MCMC framework. However, as an independent check of the stellar population parameters derived from \textsc{PyStaff}, we also measured parameters from just the strong absorption features in the optical MUSE data with \textsc{ppxf} and used different models to verify the results. An advantage of fitting the stellar populations using \textsc{ppxf} is that no prior assumptions about the SFH have to be made (in \textsc{PyStaff} a single stellar population is assumed). A useful feature of this method is that one can separate distinct populations of stars within a single galaxy spectrum based on the best-fitting templates. 

We used the E-MILES-based SSP models from \cite{Vazdekis2016} (see Section \ref{sec:derivkin}) due to their breadth and fine binning of the age and metallicity options. \textsc{ppxf} fits a weighted distribution of models to the spectra, so having a fine grid of models to choose from provides more detailed information on the stellar population parameters. We fitted the MUSE spectra between $\sim4750\text{--}5850\;\text{\AA}$ and the continuum was fitted with multiplicative polynomials (with \texttt{mdegree}~=~10) as recommended in \textsc{ppxf} for stellar-population fitting. We used models with a \cite{Salpeter1955} IMF (unimodal, $x=2.3$; based on the BaSTI isochrones) with all possible ages (53 spanning $0.03\text{--}14.00$~Gyrs) and total metallicities (12 between [M/H]~$=-2.27$ to $+0.40$). We verified that the choice of IMF in this optical region was not important and got the same results using all other available IMF templates: \cite{Kroupa2001} universal and revised, Salpeter, \cite{Chabrier2003} and bimodal \citep{Vazdekis1996}, all with a high-mass end slope of $x=2.3$ (i.e. Salpeter). The \textsc{ppxf} fits to the MUSE spectra and corresponding metallicity and age grids for the radial bins are shown in Figure \ref{fig:ppxfsps}.

In order to extract useful information about the stellar populations when spectral fitting with \textsc{ppxf}, regularization (smoothing of the template weights whilst fitting) must be used \citep{Cappellari2017}. We used a regularization parameter \texttt{regul}~=~100 which is advised when the templates are normalised. Using the fits to the MUSE spectra for the radial bins and the regularized grids of weights, we derived mass-weighted stellar population parameters for the three regions across the surface of the galaxy. 

To estimate the uncertainties on the derived stellar population parameters, we adopted a method suggested by M.~Cappellari (private communication). From a spectral fit to a galaxy spectrum, the best fit to the data and the residuals ($=\text{galaxy}-\text{best fit}$) can be used to generate bootstrapped spectra that were refitted with \textsc{ppxf}. The method employs the Wild bootstrap \citep{Wu1986} that has the following form:
\begin{equation} \label{eq:wild}
y^*_i = \hat{y}_i + \hat{\epsilon}_i v_i.
\end{equation}
Where $y^*_i$ is the bootstrapped value for wavelength pixel $i$, $\hat{y}_i$ is the best fit to the real data for pixel $i$, $\hat{\epsilon}_i$ is the residual of the best fit to the real data for pixel $i$, and $v_i$ is a random variable that can have a value of $\pm1$ with a probability of a $\frac{1}{2}$. Using this equation, we generated 100 bootstrapped spectra from the single best fit and residuals from our original fit to the data. We then fitted each with \textsc{ppxf} using the same method as before but with the velocity and dispersion derived from the data held fixed. We derived 1$\sigma$ errors from the distribution of the bootstrap spectra values and these are given in Table \ref{tab:fsstellpop} along with the derived values.

\begin{figure}
\centering
\includegraphics[width=0.49\textwidth, trim=0 10 0 0, clip]{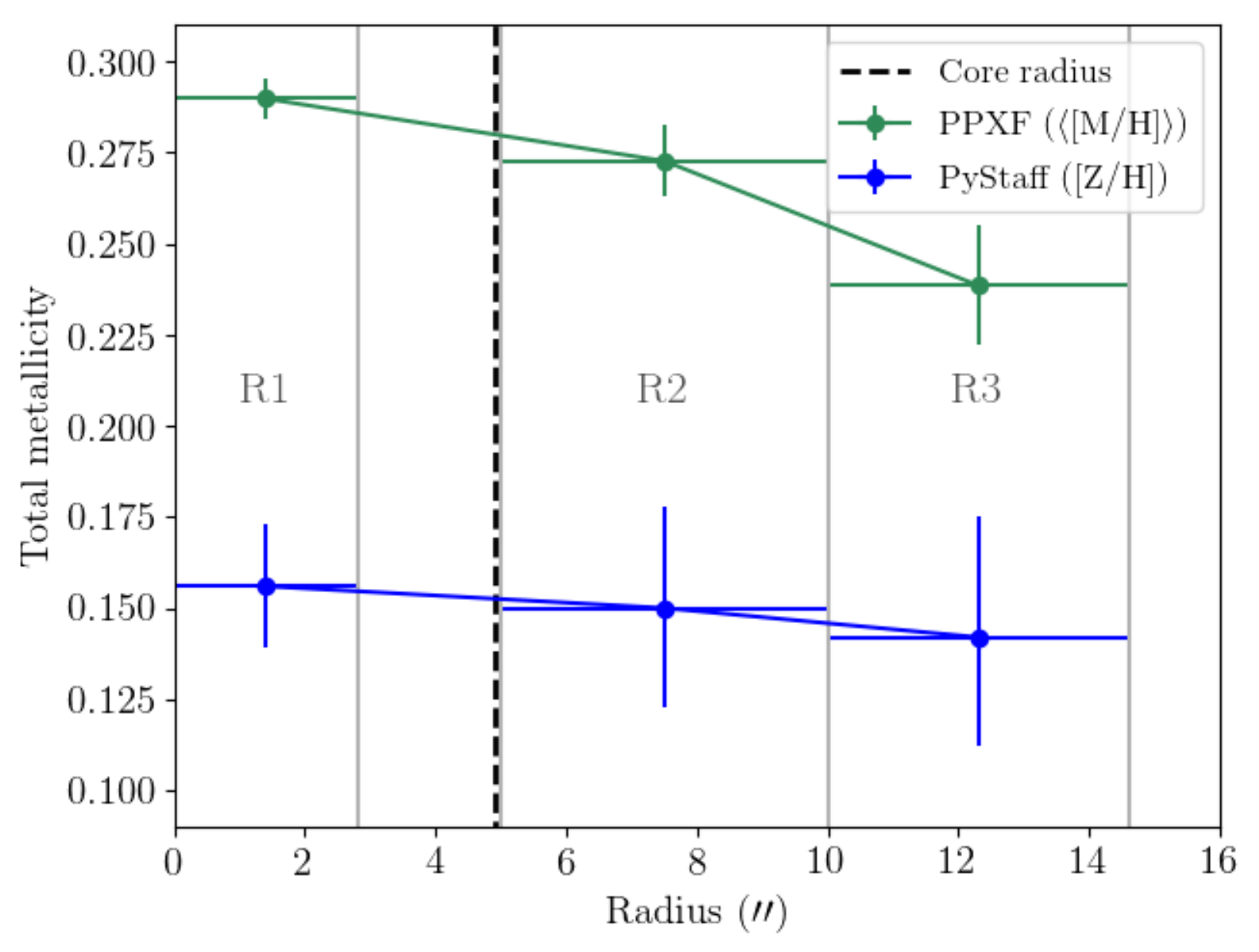}
\caption[]{Total metallicities as a function of radius for IC~1459 derived for each radial bin using the different methods of \textsc{ppxf} ($\langle$[M/H]$\rangle$; green) and \textsc{PyStaff} ([Z/H]; blue). The radial bins, core (R1), middle (R2) and outer regions (R3) of IC~1459 are shown (grey vertical lines), along with the kinematic core radius (black dashed line). The bins correspond to the concentric circles in Figures \ref{fig:musekinbin}--\ref{fig:kmossigbin}. The relative offset of the measurements comes from the differences in the fitting methods, models, and metallicity diagnostics and should not be taken as a direct comparison. We show them here to compare the trends found by the two fitting methods. \textsc{ppxf} finds a negative metallicity gradient that could be driven by a distinct low-metal population show in Figure \ref{fig:ppxfsps}. \textsc{PyStaff} assumes a single stellar population and finds a slightly negative but relatively flat metallicity gradient.}
\label{fig:metcom}
\end{figure}

Figure \ref{fig:metcom} shows the metallicity values derived with \textsc{ppxf} ($\langle$[M/H]$\rangle$, \citealt{Vazdekis2016} models) and \textsc{PyStaff} ([Z/H], \citealt{Conroy2018} models) as a function of radius to evaluate the trends derived from each. A direct comparison of the output values is advised against due to the differences in the two fitting methods (\textsc{PyStaff} assumes an SSP whereas \textsc{ppxf} fits for multiple stellar populations and is a mass-weighted measurement), fitting regions (blue part of MUSE with \textsc{ppxf}) and models. The different stellar population synthesis codes determine metallicity in different ways, both produce a `total metallicity' measurement but we do not expect these to be consistent. The radial trend from \textsc{ppxf} shows a decrease in metallicity with radius. This is also seen in \textsc{PyStaff}, although the gradient is consistent with a constant value within errors.

We found older ages and higher metallicities using \textsc{ppxf}. The ages determined using \textsc{ppxf} are constant within errors ($\sim 12$ Gyr) compared to a negative gradient and younger ages measured with \textsc{PyStaff}. In contrast, we found a negative metallicity gradient with \textsc{ppxf} and constant metallicity with \textsc{PyStaff}. In the case of \textsc{ppxf}, we found that this metallicity gradient was driven by a metal-poor population that becomes increasingly dominant with radius. We only made measurements in three bins, but it is likely that the trend changes smoothly with radius.

The fit to the outer radial bin and the grid showing this dual population are shown in bottom panels of Figure \ref{fig:ppxfsps}. The bulk of the stellar population is old and metal-rich (upper right spread in grid). The metal-poor population (spread in the bottom right of the grid) makes up $\sim16\%$ of the fitted models. This indicates that the outer region of IC~1459 (R3) hosts a dual stellar population: a metal-rich population, as seen throughout the galaxy, and a metal-poor population that becomes more dominant in the outer regions. It seems likely that the metal-poor population, superimposed on the metal-rich population, is driving the overall metallicity gradient of this galaxy. 

Such a low metallicity model has very few spectral features and so could be used by \textsc{ppxf} to help fit the continuum along with the polynomial. However, we do not detect the low-metal populations towards the centre (top and middle panels of Figure \ref{fig:ppxfsps}), so if it was purely a problem with the fit then perhaps this should be seen in every radial bin. Previous studies of IC~1459 also measure this negative metallicity gradient \citep[][]{Carollo1993, Amblard2017}. This radial trend in metallicity has also been reported for several other KDCs \citep[e.g.,][]{Efstathiou1985, Franx1988, Gorgas1990, Bender1992a, Mehlert1998, Davies2001, Emsellem2004, Kuntschner2010}.

We confirmed that this result of a dual population was not an effect of coarse binning and poorer quality spectra towards the galaxy outskirts by inspecting \textsc{ppxf} fits of the stellar populations in individual Voronoi bins using the MUSE S/N $=100$ map. Despite the slightly increased noise, the stellar-absorption features are strong in the individual bins and the fits to the data look good.

\begin{figure}
\centering
\includegraphics[width=0.5\textwidth, trim=25 25 25 25 ,clip]{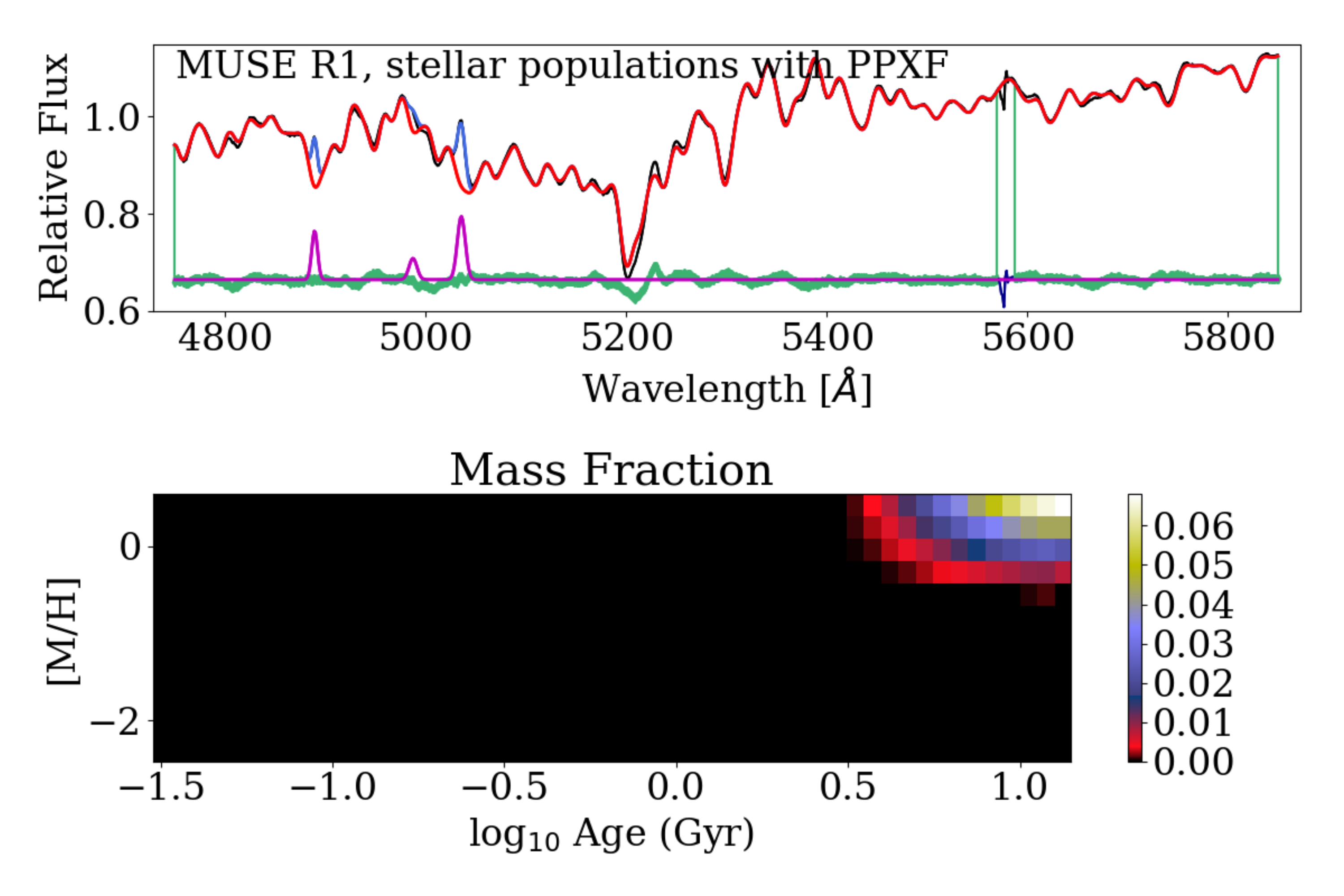}\\ \vspace{5mm}
\includegraphics[width=0.5\textwidth, trim=25 25 25 25 ,clip]{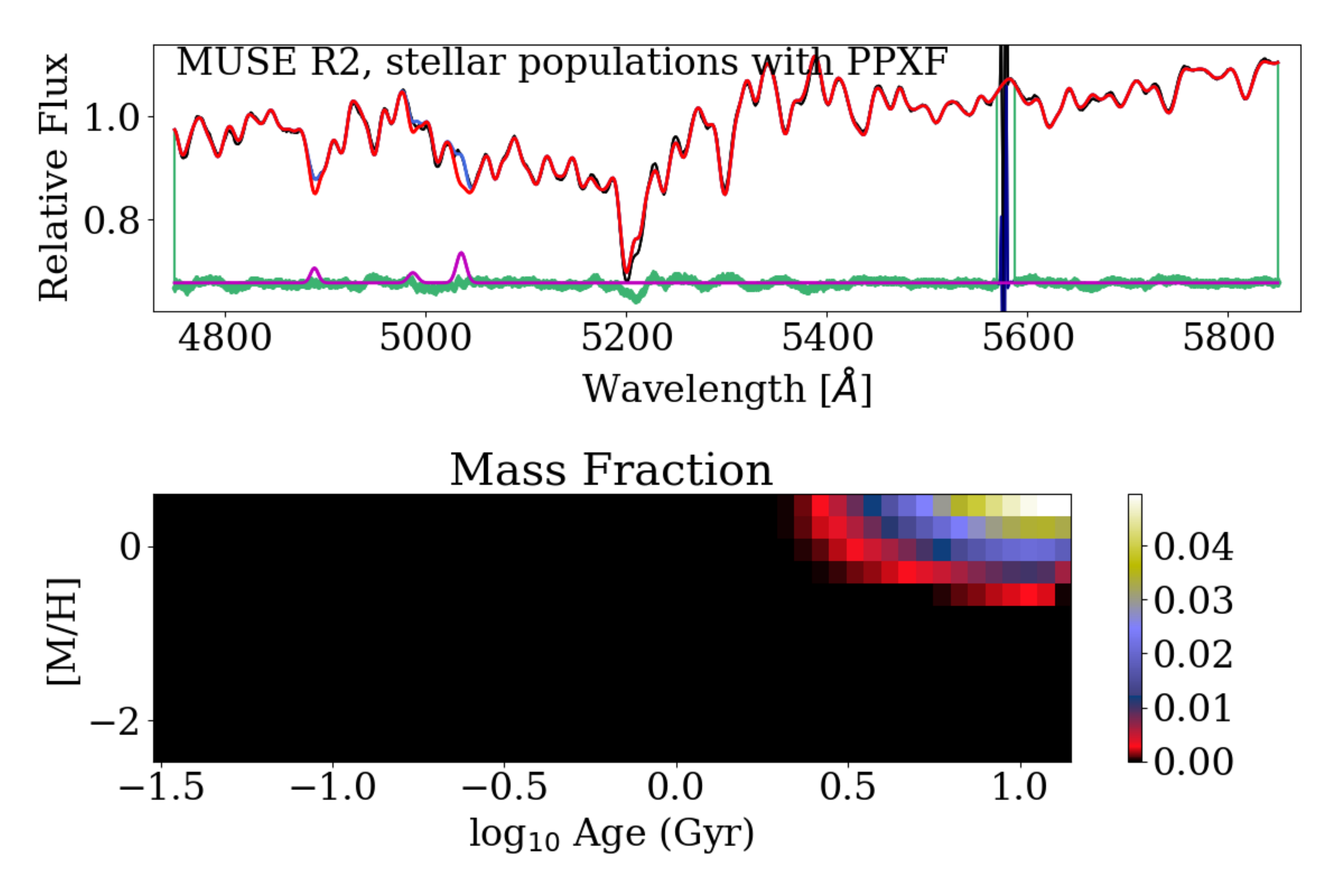}\\ \vspace{5mm}
\includegraphics[width=0.5\textwidth, trim=25 25 25 25 ,clip]{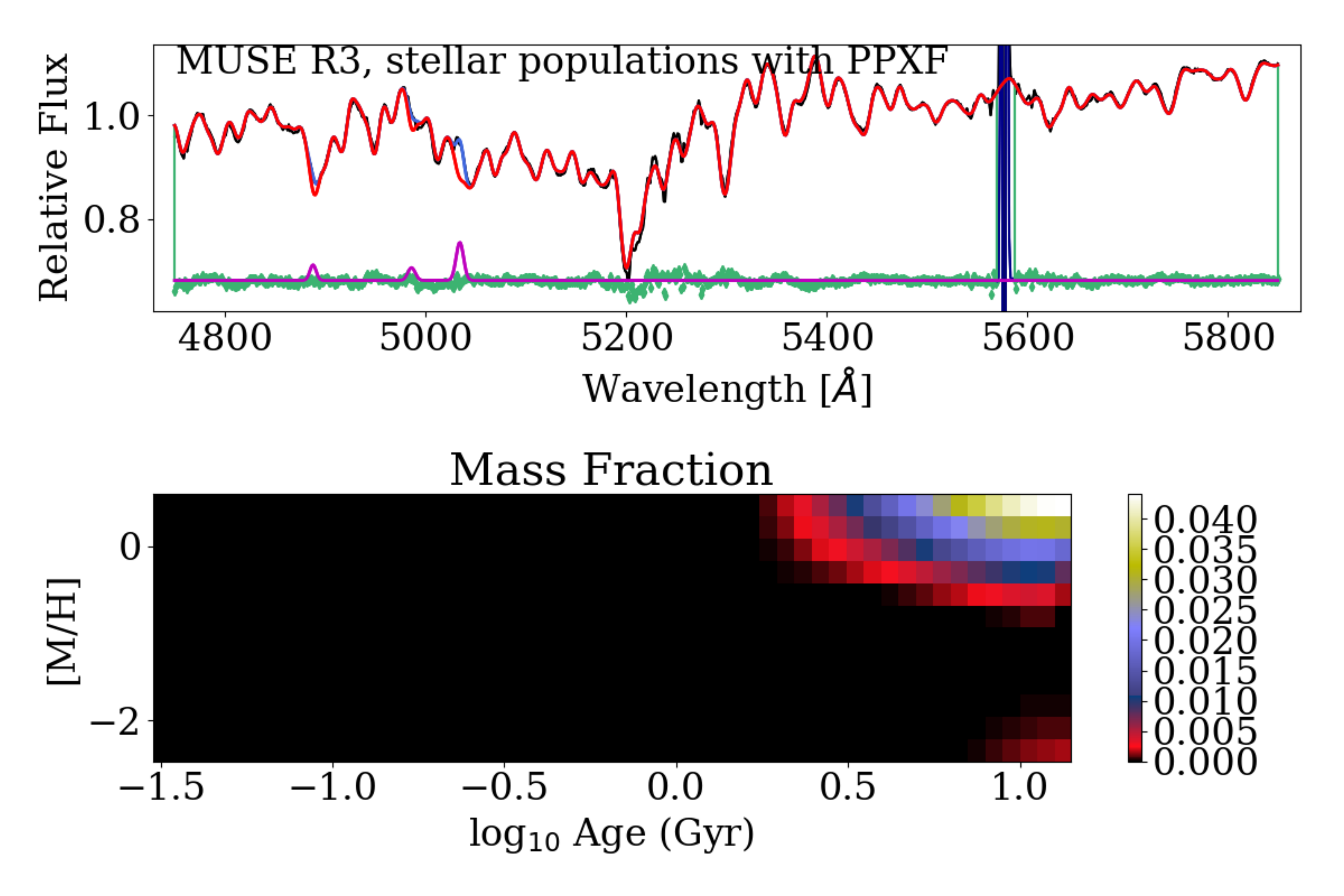}
\caption[]{\textsc{ppxf} fits to MUSE radially binned spectra of IC~1459 to extract stellar population parameters for comparison with \textsc{PyStaff}. Fits to the central (R1, top panels), middle (R2, middle panels), and outer (R3, bottom panels) regions of IC~1459 are shown. The spectra (black), the \textsc{ppxf} best fit using regularization (red), fit to the emission lines (pale blue in spectrum, magenta spectrum below), residuals (green) and masked sky line (dark blue) are shown. The grid shows the full spread of ages and metallicities of the E-MILES-based SSP models from \cite{Vazdekis2016} and are colour-coded for mass fraction. See Section \ref{sec:ppxf}.}
\label{fig:ppxfsps}
\end{figure}

\section{Discussion}
\label{sec:icdisc}

\subsection{The KDC kinematics}
\label{sec:disckin}

IC~1459 has high dispersion along its major axis (i.e. south east to north west, Figure \ref{fig:musesigbin}). This is expected for sharp velocity gradients (seen along the major axis in Figure \ref{fig:musekinbin}), where multiple components are travelling at different velocities and broadening the spectral features. The higher major-axis dispersion was also found for other massive ETGs with KDCs: NGC~5813 \citep[core counter-rotating;][]{Krajnovic2015} and NCG~4365 \citep[core rotating at $\sim45^{\circ}$;][]{Davies2001, vandenBosch2008}. NCG~5813 was also observed with MUSE but for much longer and with better seeing than IC~1459 ($\sim80$~min on source, seeing $\sim0.7^{\prime\prime}$). The dispersion map of NCG~5813 showed that the high major-axis dispersion was actually in two peaks symmetrically separated from the core. The poorer seeing of our MUSE data ($\sim2^{\prime\prime}$) may mean that these peak features are not resolved or that they are not present. We do see that the dispersion remains high in a shoulder to the north west of the centre in R2 after which it falls away (Figures \ref{fig:musesigbin}, bottom panel, and \ref{fig:sigslice}). More relevant is the dispersion map of NCG~4365 in \cite{vandenBosch2008}, that shows an hourglass shape where the dispersion along the major axis flares outward at larger radii. 
    
Another faint feature of the MUSE dispersion maps for IC~1459 is the asymmetry (dynamically hotter to the south east). This may be a feature of the reduction, perhaps relating to the sky subtraction, or a genuine kinematic feature of the galaxy; perhaps the result of a merger as is indicated by other observational evidence (Section \ref{sec:merger}). However, at these central radii, the dynamical time is relatively short so one would assume the core to be fully relaxed, which suggests the feature may not be real.

A previous kinematic study of IC~1459 used interpolation between six long-slit spectra to produce a 2D rotation map of IC~1459 \citep{Cappellari2002}. Schwarzschild modelling was then used to measure the stellar kinematics of the galaxy and understand its counter-rotating core. Results from this study suggested that the stars in the core are on distinctly different orbits to the rest of the galaxy. Since this work, advances such as IFU data (that enabled more detailed Schwarzschild modelling of other galaxies) and improved spectral fitting techniques \citep[that fit for absorption and emission simultaneously, e.g. \textsc{ppxf};][]{Cappellari2004, Cappellari2017} have allowed us to improve our understanding of KDCs.

The \cite{Schwarzschild1979, Schwarzschild1982} orbit-based dynamical modelling method was used to isolate the orbital information of just the KDC in the E3 elliptical NCG~4365 by modelling data from SAURON \citep{vandenBosch2008}. The galaxy has a constant radial age \citep{Davies2001}. \cite{vandenBosch2008} discovered that there is no distinct orbital component in NCG~4365. Instead, it was found that the galaxy was well modelled by the superposition of two counter-rotating short-axis thick tubes of orbits \citep[one of the four stable orbit families of a triaxial system, see e.g.][]{Statler1987}, each of which has a smooth distribution with radius. The $\sim45^{\circ}$ misalignment of the core and outer part of the galaxy came about from the addition of the long-axis and box orbital components that showed net rotation. A similar composite-orbit conclusion was found from modelling of NCG~5813 \citep{Krajnovic2015} using high-S/N MUSE data. 
    
To explain this picture, \cite{vandenBosch2008} proposed a galaxy that has undergone a number of randomly distributed dry mergers. The result would be an orbital distribution with no clear preferred rotation and dominated by random motions. However, near the centre where the density varies more rapidly, any small mismatch between the orbital families can result in an apparent KDC. Without full Schwarzschild modelling of high-S/N IFU data, we cannot know whether this proposed theory would also fit IC 1459. However, it could help explain some of IC 1459's observed properties, such as the potential hourglass shape seen in its dispersion. A similar feature is prominently displayed in the fourth row of Figure 13 of van den \cite{vandenBosch2008}, showing the combined prograde and retrograde short-axis orbits for NGC 4365.

\subsection{Possible formation scenarios of IC~1459}
\label{sec:discform}

There could well be numerous possible formation processes of galaxies with KDCs as is suggested by their wide range of stellar populations \citep{Bender1992a, Carollo1997a, Kuntschner2010} and kinematic properties \citep{Krajnovic2008, Krajnovic2011}. What are the possible mechanisms that could lead to the imbalance of stellar mass in the prograde and retrograde orbits that could result in a radially constant  age? The possible mechanisms would have to include either an early assembly or a late assembly of systems with similarly aged populations \citep[e.g.,][]{Davies2001, Kuntschner2010}. We outline three possible formation mechanisms and evidence that supports or conflicts with each.

\subsubsection{Two-phase assembly}
\label{sec:2phase}

In the `two-phase' evolution scenario \citep[e.g.,][]{Oser2010}, the central regions of ETGs form in a rapid central starburst event which results in a dense stellar core. Their outer layers are then accreted over time in a series of dry minor mergers. There is now significant observational evidence for extreme, early star formation at $z\gtrsim2$ that fits with the first phase \citep[e.g.,][]{Bezanson2009, Barro2013, Nelson2014, Prichard2017b}. This could result in a  bottom-heavy central IMF \citep[e.g.,][]{Chabrier2014,vanDokkum2015, Zolotov2015, Barro2016, Kriek2016}. 

Radial measurements of IMF-sensitive absorption features in some massive ETGs have inferred an increasingly bottom-heavy IMF towards their centres \citep[e.g.][]{Martin-Navarro2015a, Martin-Navarro2015b, LaBarbera2016, Conroy2017, vanDokkum2017, Parikh2018, Sarzi2018, Vaughan2018b}, although there are differences in the precise shape of the radial IMF profiles. However, other studies do not see this effect, measuring little radial variation \citep[][]{Alton2018} or a MW-like IMF \citep{Zieleniewski2017}. Furthermore, a number of studies have found MW-like IMF normalisations from gravitational lensing \citep[e.g.,][]{Smith2013, Smith2015a, Smith2015b, Collier2018a, Collier2018b}. Some of these differences can be attributed to aperture differences between studies \citep{vanDokkum2017} or other systematics between methods used to measure the IMF, but the tensions between techniques is still not fully understood \citep[e.g.,][]{Smith2014, Smith2015a, Newman2017}. Some scatter also occurs within uniformly analysed samples of galaxies \citep[][]{Conroy2012b, Leier2016, vanDokkum2017}, implying the disparities are likely to be caused by the effects of different evolutionary paths.

The photometric profile and metallicity gradient of IC~1459 adds weight to its formation via a two-phase evolutionary path \citep[e.g.,][]{Oser2010} that has undergone accretion by smaller gas-poor satellites at later stages of formation \citep[e.g.,][]{Naab2007, Naab2009, Hopkins2009}. IC~1459 has a high S\'{e}rsic index \citep[$n\sim8.25$ when masking the core;][]{Lasker2014} which is indicative of intermediate to minor mergers \citep[with mass ratios 1:5 to 1:10;][]{Hilz2013}. The presence of a metal-poor population measured with \textsc{ppxf} in the outer regions of the galaxy also adds weight to this second phase, where the outer layers could be accreted lower-mass and metallicity satellites. The second dry merger phase also fits with the theory proposed by \cite{vandenBosch2008} for the formation of a KDC, although this does not include a first phase of rapid star-formation to create the core.

The IMF results from \textsc{PyStaff} show a roughly radially constant IMF `mismatch' parameter ($\alpha$) of $\sim$2.0 to $\sim\nicefrac{1}{3}R_{\rm e}$, the limit of the outer radial bin (R3). The average profile of the six galaxies analysed in \cite{vanDokkum2017} had a bottom-heavy IMF ($\alpha\sim2.5$) in the central regions, followed by a rapid flattening with galactocentric radius to be MW-like at $\gtrsim0.4R_{\rm e}$. This trend is generally supported by further observational evidence from other galaxies \citep[e.g.,][]{Sarzi2018}, but some studies find a more extended bottom-heavy central IMF profile \cite[e.g.,][]{Vaughan2018b}. NGC 1399, the brightest early-type galaxy in the Fornax cluster and a E0--E1 cD galaxy \citep{Rusli2013}, shows a similar trend in the IMF radial profile to IC~1459 with $\alpha=1.97\substack{+0.19\\-0.16}$ that is flat to $0.7R_{\rm e}$ \citep{Vaughan2018b}.
    
IC~1459 is a `core' galaxy \citep{Lasker2014}, i.e. it has a depletion of light towards its centre \citep[e.g.,][]{Ferrarese1994, Lauer1995, Ferrarese2006}. This depleted photometric core profile of IC~1459 is typical for massive slow-rotator galaxies and is found in almost all of the slow-rotating galaxy population \citep[e.g.,][]{Emsellem2011, Lauer2012, Krajnovic2013}. The creation of the core galaxies is thought to require binary super-massive black hole (SMBH) interaction in a dissipationless merger to remove some of the stars \citep{Begelman1980, Milosavljevic2001}. A major merger may explain how a dense stellar core formed in the first phase could be flattened due to the merger of SMBHs. This may also have an effect on the bottom-heavy nature of the IMF, that could potentially be ``diluted'' by a regular less bottom-heavy IMF. The observations do not rule out the two-phase theory for IC 1459, however, there is other evidence that suggests alternative routes or ``phases'' of its evolution beyond the classical two-phase model.

\subsubsection{Merging spirals}
\label{sec:merger}

The formation of KDCs has long been thought to involve accretion from a gas-rich companion or major mergers of gas-rich spirals \citep[e.g.][]{Bender1992a, Surma1995, Hoffman2010, Bois2011, Moody2014, Tsatsi2015}. Based on the unusual kinematics of NGC 5813, \cite{Krajnovic2015} argued for the similarity with the much less massive \citep[ $ < 5 \times10^{10}\,\mathrm{M}_{\odot}$;][]{Cappellari2013a}, fast-rotating 2$\sigma$ galaxies whose kinematics display similar symmetrically offset dispersion peaks \citep{Krajnovic2011}. 2$\sigma$ galaxies are the superposition of two discs as verified by studying their stellar populations \citep{Coccato2011, Coccato2013, Johnston2013} and from modelling \citep{Cappellari2007}. They are thought to form from the accretion of gas onto a disc but with opposite angular momentum. This could occur through an interacting pair of galaxies \citep{Vergani2007, Coccato2011}.

IC~1459 is a convincing case for a galaxy that has undergone at least one gas-rich major merger. It has a high gas content and shows both dust features \citep{Sparks1985} and ionised gas emission (\citealt{Phillips1986a}; \citetalias{Franx1988}) in its core. A deep image of IC~1459 revealed tidal features in the galaxy \citep{Malin1985}, while \cite{Forbes1995} found shell-like remnants. The long tidal tails, high gas content and shells are indicative of two merging spirals \citep[e.g.,][]{Toomre1972, Toomre1977, Malin1980}. \cite{Saponara2018} found large HI clouds near IC~1459, concluding these were likely to be debris from tidal interactions with neighbouring galaxies. Being the central galaxy in a relatively small group would mean a high probability of mergers due to the low dispersion of the group \citep{Ostriker1980, Tremaine1981}. Comparing dynamical models derived from long-slit spectra of IC~1459 to simulations \citep{Bendo2000} revealed the outer parts of the galaxy were well matched to a merger of disc galaxies with a 3:1 mass ratio \citep{Samurovic2005}. The asymmetry in the dispersion map towards the outskirts of IC~1459 may be a sign of tidal interactions or even a major merger. 

Despite the evidence pointing towards a scenario of merging spirals it is unlikely that this is the sole mode of formation for galaxies with KDCs. There is conflicting evidence that ETGs are more compact than spirals and merger remnants are less compact, and velocity dispersions are higher and scale lengths are longer in ETGs than discs \citep{Ostriker1980}. In addition, the photometric profile of IC~1459 is more similar to ETGs thought to have undergone a two-phase evolution, and not the merger of spirals.

A galaxy that exhibits similar radial IMF trends to IC~1459 is NGC 1600, with roughly constant $\alpha\sim2.0$ at $\lesssim0.4R_{\rm e}$ \citep[][]{vanDokkum2017}. NGC 1600 is an isolated E3--E4 ETG with an extensive satellite galaxy population \citep[][]{Smith2008}. It is thought to be the result of a loose group of galaxies that merged $\sim4$ Gyrs ago. This is similar to the current state of the IC~1459 group, where IC~1459 is the central galaxy of a spiral-dominated group that shows clear observational evidence of spirals having merged with it. However, the bottom-heavy IMF indicates a different evolutionary path for the central galaxy prior to the group merging scenario. This may not fit with the two-phase path though, as the IMF is not as bottom-heavy as some cores measured ($\alpha\gtrsim2.5$). The full-spectral-fitting methods of \cite{vanDokkum2017}, \cite{Vaughan2018b}, and this work are comparable, and the same SSP models \citep{Conroy2018} were fitted in each case. This makes any differences in the IMF profiles a more direct result of differing galaxy formation properties rather than systematic differences between the studies. The difference between the three galaxies with similar radial-IMF profiles (group central IC~1459, isolated NGC 1600 and brightest cluster galaxy NGC~1399), makes drawing solid conclusions on a single evolutionary path for this $\alpha$ value and trend challenging. 

\subsubsection{Cold-stream accretion}
\label{sec:stream}

The counter-rotation of cold streams in the early Universe has also been proposed as a potential evolutionary path of galaxies with KDCs \citep{Keres2005, Keres2009, Dekel2009a, Krajnovic2015}. The counter-rotating populations could be explained by two cold streams that supply gas to the galaxy and exert torques to produce counter-rotating discs \citep{Algorry2014}. However, this formation scenario is gas rich which makes the depleted stellar core profile of IC~1459 (and NGC 5813) hard to explain.

We measured a relatively constant metallicity gradient in IC~1459 with \textsc{PyStaff} but a negative gradient with \textsc{ppxf} \citep[as determined in previous studies of IC~1459 e.g.,][]{Carollo1993, Amblard2017}. This radial trend in metallicity has also been reported for several other KDCs (e.g., \citealt{Efstathiou1985}; \citetalias{Franx1988}; \citealt{Gorgas1990, Bender1992a, Mehlert1998, Davies2001, Emsellem2004, Kuntschner2010}). The fitting of the MUSE data with \textsc{ppxf} revealed that this may be being driven by a distinct metal-poor population of stars in its outskirts. The low-metal population is coincident with where the retrograde motion of the outer parts of the galaxy becomes visible ($R>10\arcsec$). The result could imply that the retrograde stellar population has a lower-metallicity component that drives the drop in metallicity.

From fitting the MUSE spectra with no prior assumption of SFH with \textsc{ppxf}, we found roughly radially constant and older ages of $\sim12.0$--$11.1$~Gyrs. Interestingly, this radially constant old stellar population has been measured for several KDCs (e.g., \citetalias{Franx1988}, \citealt{Bender1988, Rix1992, Davies2001, McDermid2006b, Kuntschner2010, McDermid2015, Krajnovic2015}). The radially constant age and IMF support findings from orbital modelling that the KDC is actually co-spatial counter-rotating populations. The bottom-heavy nature of the IMF could mean there was a more extreme mode of star formation than we see locally, but less so than the starburst mode proposed in the two-phase evolution. Star formation resulting from the counter-rotating gas streams at early times could explain these stellar population results.

\section{Conclusions}
\label{sec:icconc}

We investigated IC~1459, a massive local E3 ETG with a rapidly counter-rotating core. This galaxy is part of the more general class of massive slow-rotator galaxies that have KDCs. We coupled IFU data in the optical from MUSE with KMOS data in the NIR to study the spatially resolved kinematics, stellar populations and IMF in order to understand its evolution. We used \textsc{PyStaff}, an MCMC full-spectral-fitting routine and state-of-the art SSP models \citep{Conroy2018}, to fit the stellar population and IMF parameters simultaneously, in line with other studies of radial IMF measurements. To compare to \textsc{PyStaff}, we also fitted the age and metallicity from the strong absorption features in the optical MUSE data with \textsc{ppxf}, which does not assume a SFH (although varying the IMF simultaneously is not supported). The key findings of this work are summarised below.

\begin{enumerate}
\item The velocity dispersion maps show that IC~1459 is dynamically hot along the major axis, as found for other IFU studies of KDCs \citep{vandenBosch2008, Krajnovic2015}. Unlike other studies we see asymmetry in the velocity dispersion map which may be the result of a major merger or tidal interaction as evidenced by other observations or an issue with the data reduction. There is a hint of an hourglass shape in the dispersion maps that is similar to that found by \cite{vandenBosch2008} who, from full orbital modelling of another galaxy, concluded that the KDC was a projection of smoothly distributed co-spatial counter-rotating populations.
\item We measured a negative age gradient and constant metallicity for IC~1459 with \textsc{PyStaff}. In contrast, we found a constant radial age and negative metallicity gradient with \textsc{ppxf}. The lack of an age gradient across the galaxy from \textsc{ppxf} is consistent with other studies of KDCs. The negative metallicity gradient found most strongly with \textsc{ppxf} for IC~1459 is also seen in other galaxies with KDCs. However, for the first time we have been able to identify (with \textsc{ppxf}) that the metallicity gradient may actually be being driven by an extremely metal-poor population that is more dominant in the outer parts.
\item We made radially resolved measurements of the IMF in IC~1459 using \textsc{PyStaff}; the first time this has been done for a galaxy with a KDC. For all three radial bins, we found a constant and bottom-heavy IMF with $\alpha\sim2.0$ to $\sim\nicefrac{1}{3}R_{\rm e}$ (the radial extent of our data) consistent with other studies. This supports a different and potentially more extreme star formation at early times when the core was formed. However, it is not as extreme ($\alpha>2.5$) as some ETGs.
\item The radially flat IMF measurements, and ages (from \textsc{ppxf}) extend across the galaxy and beyond the strong counter-rotation seen in the core. These results add weight to the theory from orbital modelling of IFU data of other galaxies that the KDC is not a distinct feature of the galaxy but is in fact the superposition of two co-extensive prograde and retrograde smooth populations of stars on short-axis tube orbits. The counter-rotation would therefore be an observational effect of a slight mass imbalance between these populations where the strongly rotating core is visible (radii $\lesssim5\arcsec$). However, without full orbital modelling of high-S/N IFU data of IC 1459, we are unable to say conclusively if this will hold for this galaxy as well.
\end{enumerate}

IC~1459 deviates in some ways from massive ETGs thought to have undergone a classical two-phase evolution \citep[e.g.,][]{Oser2010}. The results presented here of IC~1459 do not point to one clear picture of how it formed that explains the counter-rotating stellar populations along with its other observed properties. However, it could involve the accretion of counter-rotating cold gas streams in the early Universe followed by enhanced star-formation akin to the first phase of the two-phase assembly theory, and dry and wet mergers through to the present day, all of which there is evidence to support. This work emphasises the complexity and diversity in possible evolutionary paths of ETGs and our ignorance of what exactly these may be. It is evident that further studies of similar systems would help to build a more complete picture of the formation of KDCs. Specifically, we highlight that efforts to measure the spatially resolved IMF could prove interesting for this subset of massive slow rotators. We emphasise the synergy between KMOS and MUSE for this work due to their combined unparalleled spatial and spectral coverage.

\section*{Acknowledgements}
\noindent The authors thank the anonymous referee for valuable insights and comments that improved the paper. LJP thanks Dr. Joshua Warren for the cropped and additionally sky-subtracted MUSE cube of IC~1459. LJP also thanks Prof. Michele Cappellari for helpful discussions and Profs. Roberto Maiolino and Matt Jarvis for useful comments on the work. LJP was supported by a Hintze Scholarship for the majority of this work, awarded by the Oxford Hintze Centre for Astrophysical Surveys, which is funded through generous support from the Hintze Family Charitable Foundation. SPV is supported by a doctoral studentship supported by STFC grant ST/N504233/1. RLD acknowledges travel and computer grants from Christ Church, Oxford and support from the Oxford Hintze Centre for Astrophysical Surveys, which is funded by the Hintze Family Charitable Foundation.

\bibliography{ic1459_bib_cp}
%

\end{document}